\documentclass[superscriptaddress,onecolumn,floatfix,notitlepage,citeautoscript]{revtex4-2}
\usepackage[T2A]{fontenc}
\usepackage{graphicx}
\usepackage{amsmath}
\usepackage{amssymb}
\usepackage[labelfont=bf,format=plain,justification=centerlast]{caption}
\usepackage[usenames,dvipsnames]{xcolor}
\usepackage{hyperref}
\hypersetup{colorlinks=true,linkcolor=blue,urlcolor=black,citecolor=blue}
\usepackage{soul}
\usepackage{indentfirst}

\setcitestyle{super}
\usepackage{ulem}
\usepackage{lineno}
\usepackage{titlesec}
\titleformat{\section}
  {\normalfont\scshape\bf}{\thesection}{}{}
\titleformat{\subsection}
  {\small\bf}{\thesubsection}{1em}{}

\newcommand{\vor}[1]{{\color{black}#1}}


\begin{document}
\title{Exciton-polariton ring Josephson junction}

\author{Nina~Voronova}
%\email{neenoune@gmail.com}
\affiliation{National Research Nuclear University MEPhI (Moscow Engineering Physics Institute), Kashirskoe shosse 31, 115409 Moscow, Russia}
\affiliation{Russian Quantum Center, Skolkovo IC, Bolshoy boulevard 30 bld. 1, 121205 Moscow, Russia}

\author{Anna~Grudinina}
\affiliation{National Research Nuclear University MEPhI (Moscow Engineering Physics Institute), Kashirskoe shosse 31, 115409 Moscow, Russia}
\affiliation{Russian Quantum Center, Skolkovo IC, Bolshoy boulevard 30 bld. 1, 121205 Moscow, Russia}

\author{Riccardo Panico}
\email[Current affiliation: Institut f\"{u}r Angewandte Physik, Universit\"{a}t Bonn, Wegelerstra\ss e 8, 53115 Bonn, Germany]{}
\affiliation{CNR Nanotec, Institute of Nanotechnology, via Monteroni, 73100, Lecce, Italy}

\author{Dimitris~Trypogeorgos}
\affiliation{CNR Nanotec, Institute of Nanotechnology, via Monteroni, 73100, Lecce, Italy}

\author{Milena~De Giorgi}
\affiliation{CNR Nanotec, Institute of Nanotechnology, via Monteroni, 73100, Lecce, Italy}

\author{Kirk~Baldwin}
\affiliation{PRISM, Princeton Institute for the Science and Technology of Materials, Princeton University, Princeton, New Jersey 08540, USA}

\author{Loren~Pfeiffer}
\affiliation{PRISM, Princeton Institute for the Science and Technology of Materials, Princeton University, Princeton, New Jersey 08540, USA}

\author{Daniele~Sanvitto}
\email{daniele.sanvitto@nanotec.cnr.it}
\affiliation{CNR Nanotec, Institute of Nanotechnology, via Monteroni, 73100, Lecce, Italy}

\author{Dario~Ballarini}
%\email{dario.ballarini@nanotec.cnr.it}
\affiliation{CNR Nanotec, Institute of Nanotechnology, via Monteroni, 73100, Lecce, Italy}

\begin{abstract}
{\bf Abstract}.
Macroscopic coherence in quantum fluids allows the observation of interference effects in their wavefunctions, and enables applications such as superconducting quantum interference devices based on Josephson tunneling. The Josephson effect manifests in both fermionic and bosonic systems, and has been well studied in superfluid helium and atomic Bose-Einstein condensates. In exciton-polariton condensates–that offer a path to integrated semiconductor platforms–creating weak links in ring geometries has so far remained challenging. In this work, we realize a Josephson junction in a polariton ring condensate. Using optical control of the barrier, we induce net circulation around the ring and demonstrate both superfluid-hydrodynamic and the Josephson regime characterized by a sinusoidal tunneling current. Our theory in terms of the free-energy landscapes explains the appearance of these regimes using experimental values. These results show that weak links in ring condensates can be explored in optical integrated circuits and hold potential for room-temperature applications.
\end{abstract}

\maketitle

\vspace{-20pt}
\section*{INTRODUCTION}
\vspace{-10pt}
Soon after the original proposal of Josephson in 1962~\cite{josephson}, it became clear that the Josephson effects are typical not only to tunnel junctions~\cite{anderson,shapiro}, but to many other types of so-called weak links. Those are the areas of a system (superconducting or superfluid) with the size of the order of the healing length, where the system's order parameter is substantially suppressed~\cite{RMP79,gati}. The actual sizes of weak links in various systems may drastically vary, from a few~nm in the case of a superconducting tunnel junction~\cite{shapiro} up to 10--100~nm for Dayem bridges~\cite{Dayem1}, and even microns in the case of a normal-metal barrier~\cite{SNS1}.
Another example of a system exhibiting Josephson tunneling is liquid He flowing through an array of microapertures~\cite{He3RMP,He42001}. In such a system, if a pressure is imposed on one side, counter-intuitively the superflow ceases to show mass transport and starts to oscillate. If the healing length drops as the temperature is lowered much below critical, the oscillatory sinusoidal current-phase relation connecting the two sides of the weak link gives way to a linear dependence~\cite{He3,He4}. In atomic systems, the Josephson phenomena have been extensively studied for double-well potential traps, where a barrier separating the two Bose-Einstein condensates (BECs)~\cite{Javanainen,Zapata,Giovanazzi,Albiez,levy,leblanc,farolfi} provides the mechanism of coupling their order parameters. A similar setting has been realised in ultracold Fermi gas mixtures close to a Feshbach resonance, which unite Bose-Einstein condensation of tightly bound molecules and Bardeen-Cooper-Schrieffer superfluidity of long-range fermion pairs~\cite{roati_science}.

It has been argued, however, that coherent oscillations in double-well-confined BECs may be regarded as the a.c. Josephson effect and such a setting can be reasonably called a Josephson junction only in the case when there is no confinement by the two sides of the barrier, i.e. in the limit of an infinite system~\cite{pra81}, or when the barrier is high enough to suppress the order parameter to zero~\cite{leblanc}. The closest approximation of such a system can be realised by connecting the two wells in a ring geometry.
 Ring-shaped and toroidal atomic BECs with a weak link have been considered~\cite{buchler,didier, solenov}, with the main focus on the investigation of the interplay of quantum fluctuations and tunneling of bosons through the junction. Atomic-BEC rings with a rotating weak link have been shown to exhibit quantum phase slips between quantized persistent current states~\cite{campbell,roati_prx12,Hadzibabic} including the hysteresis between the neighbouring linear current-phase branches~\cite{campbell1}. In this case, due to the relatively large size of the barrier, only the hydrodynamic regime of the weak link has been reached~\cite{campbell3}. Recently, finite-circulation states in a tunable array of tunneling weak links in a ring-shaped superfluid have been realised to reveal the effect of several junctions on the stability of the atomic flow~\cite{roati24}.

Another system that has come into focus for bosonic Josephson effects and realisation of supercurrents is that of exciton-polaritons, solid-state quasiparticles stemming from the strong coupling of excitons in a semiconductor and photons in a microcavity~\cite{microcavities}. In polariton systems, the observation of oscillations (akin to a.c. Josephson effect) due to an imbalance of population between double-well-connected polariton BECs was observed under pulsed excitation~\cite{klagoud,abbarchi}. Still, no characteristic current-phase relation has been reported, and, for a barrier size comparable with the system size, this interconversion of the two BECs can more simply be interpreted as nonlinear Rabi oscillations in a coherent two-level system. At the same time, based on this progress and the mathematical analogy with superconducting flux qubits, split-ring polariton condensates have been proposed~\cite{kavokin1,savvidis}.
The nontrivial topology of the ring-shaped BECs together with the uniformity of the potential along the ring circumference makes them an ideal system for investigating persistent currents and superfluid properties of polaritons. However, to date the experimental realisations of circulating polariton currents~\cite{assmann,kavokin2,lukoshkin,snoke} were only achieved in the settings with a broken chiral symmetry without the study of a possible weak-link behavior. Compared to atomic BECs where different regimes of a weak link in a ring can be studied by rotation of the cloud~\cite{roati_prx12}---or, alternatively, rotation of the barrier,~\cite{campbell,campbell1,campbell3}---in solid-state condensates, such as exciton-polaritons, rotation of the particle ensemble or the weak link around a ring is hard to achieve due to the extremely short lifetime of such condensates. Recently, however, ultrafast temporal modulation techinques enabled for the striking demonstration of optical stirring of polariton condensates and for observation of Floquet–Bloch bands~\cite{fraser2023,bucket,fraser2024}.

Here, we demonstrate the first all-optical realisation of a ring Josephson junction within a solid-state platform of exciton-polaritons. The boundary conditions on the weak link and the supercurrent flowing through it are defined by obstructing one part of the ring from an externally imposed flow of particles otherwise present in both branches of the same ring.
In this way we can exert an effective circulation akin to a moving barrier in a static ring. % \cite{campbell3}.
In our experiments, as the barrier is changed from a vanishingly thin to a thick one, the phase difference $\Delta\phi$ across the weak link changes, and so does the velocity of superflow under the barrier $v$ which is proportional to this phase difference. For $\Delta\phi$ lower than $\pi$, $v$ is lower than the sound velocity $c_{\rm s}$ and the fluid behaves as a superfluid, with a continuously connected phase across the barrier. Phase slips leading to the appearance of a winding around the ring in this regime cannot happen as this would increase the system's energy. However, for circulations that bring $v$ close to $c_{\rm s}$, the phase difference across the barrier approaches $\pi$ and it is energetically more favourable to reduce the amplitude of the order parameter within the barrier. At critical circulations, a true Josephson junction is formed with zero density across the barrier and phase discontinuity at its extremes. This state will be described in our theoretical model with the appearance of two degenerate minima in the system's free energy landscape. This leads to two possible outcomes of the same-setting experiment: %, one with and one without a vortex state.
one with zero and one with non-zero quantized circulation of phase gradient around the ring. As a result, when experiments are repeated many times, supercurrent states created in many realisations form statistics of zero and non-zero windings of phase.
These are key conditions for establishing of the Josephson regime. Here, a sinusoidal behaviour of the tunneling current versus applied circulation is demonstrated while the hydrodynamic flow is suppressed. For even larger barriers, $\Delta\phi$ exceeds $\pi$ and the order parameter in the region of the weak link is forced to reconnect again, recovering the hydrodynamic regime with velocity circulation increased by one quantum number.
Our experiments provide direct evidence of the switching between hydrodynamic and Josephson regimes in an exciton-polaritons quantum fluid.

\begin{figure}[b]
\centering
\includegraphics[width=\linewidth]{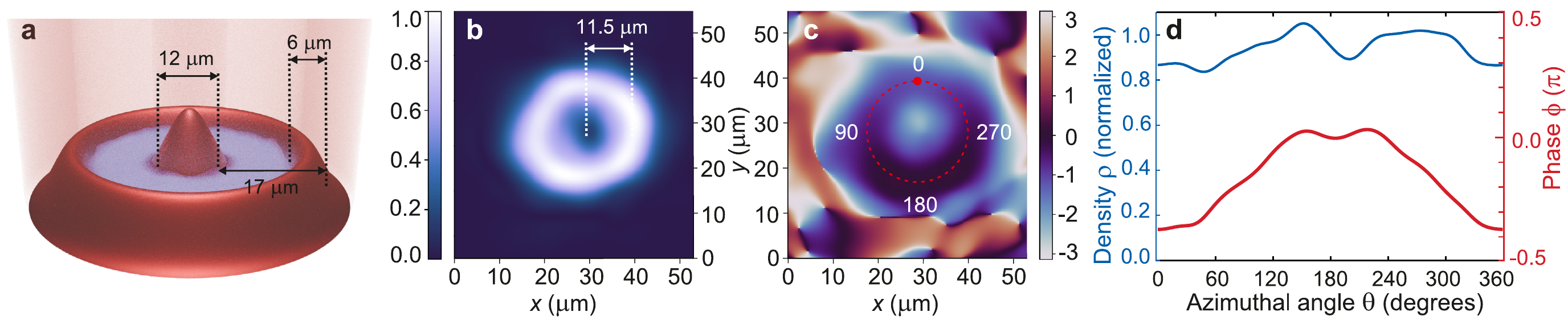}
\linespread{1.0} \caption{{\bf Polariton ring with an imposed flow.} (a) Schematic of the non-resonant pump profile (red) creating the potential confining the polariton BEC (blue). Dimensions of the pump are given on the panel. (b) Experimental steady-state PL image of a ring-shaped polariton condensate in the uninterrupted case, and (c) the corresponding phase distribution dictated by the microcavity wedge. (d) Experimental density (blue) and phase (red) profiles of the annular polariton BEC with imposed flow, taken along the red dashed line in panel {\bf c}, counter-clockwise with the zero of azimuthal angle chosen at the top of the ring (bold red point). While the density is almost uniform, the linear slopes of the phase along the two halves of the ring reveal two supercurrents flowing in opposite directions, that results in a total circulation equal to zero.
}
\label{fig1}
\end{figure}

\vspace{-10pt}
\section*{RESULTS}
\begin{figure}[b]
\centering
\includegraphics[width=0.7\textwidth]{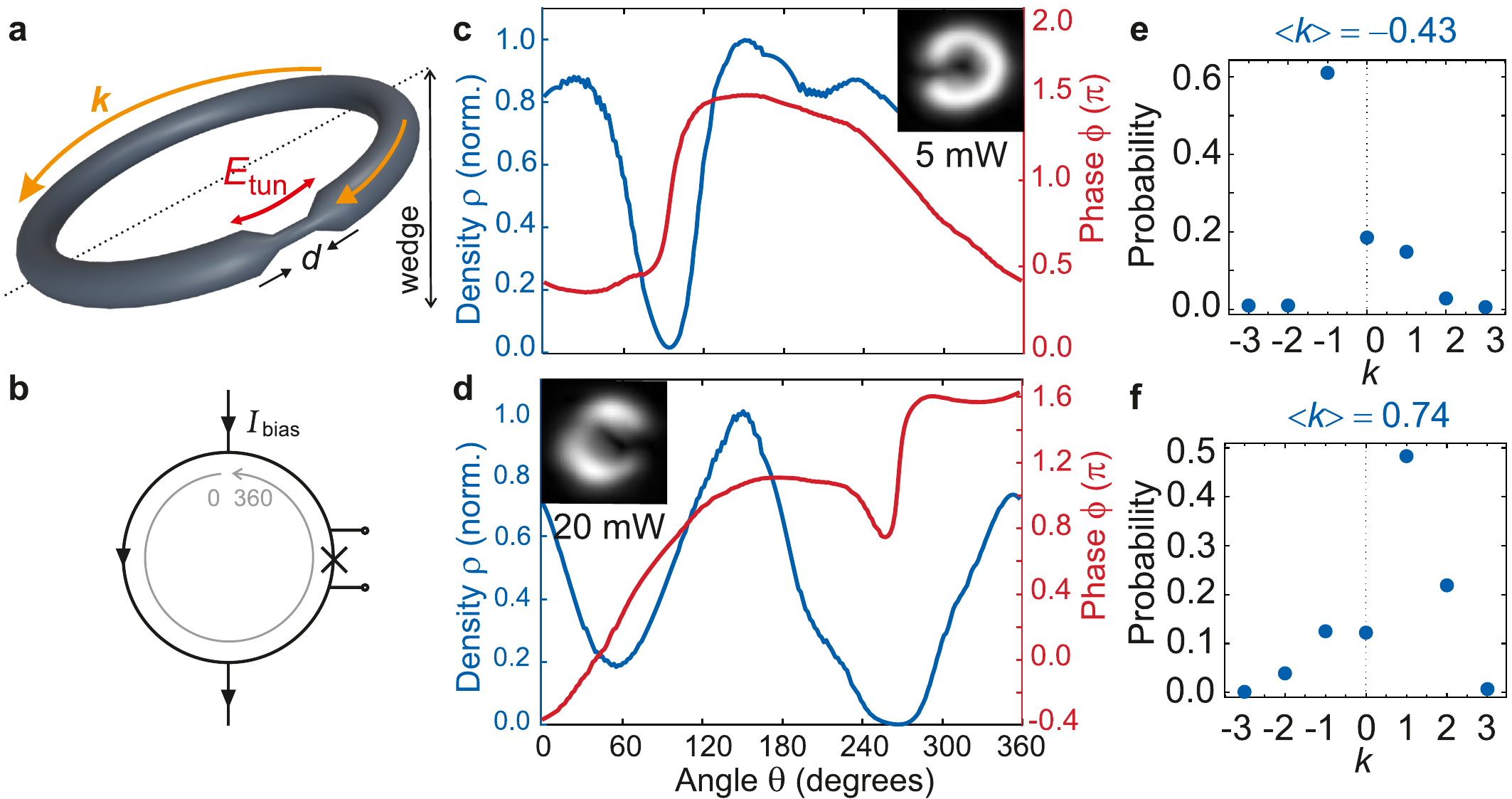}
\caption{\small\linespread{0.8} {\bf Controllable imposed circulation along the tilted polariton ring-shaped BEC.}
(a)  Sketch of the tilted (due to the microcavity wedge) polariton ring interrupted on the right-hand side by a weak link of effective width $d$. The yellow lines show the direction of the flow ${\bf k}$. The red arrow marked $E_{\rm tun}$ illustrates the tunneling energy (see text).
(b) An electric-circuit analog of the experiment: a current-biased ring with the Josephson junction placed on the right arm. The black arrows indicate the direction of the current $I_{\rm bias}$. The gray circular arrow indicates the azimuthal angle axis.
(c--d) Exemplary density (blue) and phase (red) profiles along the central radius of the ring, for the barrier of power 5~mW placed on the left side (c) and the barrier of power 20~mW placed on the right side (d) of the ring. The insets show the experimental 2D intensity profiles for these cases. (e--f) The corresponding distributions of $k$ according to Eq.~(\ref{<k>}) and the average imposed circulations (given on top of the panels) provided by the chosen barriers. By choosing the position of the barrier (left or right) and varying the power from 3~mW to 20~mW for each position we span a wide range of $\langle k\rangle$ from negative to positive values.
}
\label{fig2}
\end{figure}

\vspace{-12pt}
\noindent\textbf{Annular polariton BEC with imposed circulation.} Exciton-polaritons present a powerful platform to study collective phenomena in the solid state. They are endowed with interactions inherited from their exciton component and, at the same time, they have a light effective mass and high degree of spatial and temporal coherence thanks to the underlying photons~\cite{QFL}. Furthermore, the driven-dissipative nature of polariton systems leads to the high accessibility and control over both the excitation and readout, allowing for optical shaping of such superfluids into desired configurations~\cite{sanvitto_nphot,QFL}.
Here, as shown in Fig.~\ref{fig1}{\bf a}, we form the annular polariton condensate by the continuous-wave pumping laser of a Mexican-hat profile that provides conditions for polaritons to reach the threshold density at the bottom of the potential. The polariton condensate is coherent along the whole ring, allowing the measurements of density (Fig.~\ref{fig1}{\bf b}) and phase (Fig.~\ref{fig1}{\bf c}) with interferometric techniques (see Methods for details). The phase profile shown in Fig.~\ref{fig1}{\bf d} highlights the phase gradient, which corresponds to a finite velocity from the top to the bottom of the ring. This is caused by a wedge in the cavity thickness (see the Supplementary Information, SI), accelerating polaritons in the direction of the wedge~\cite{snoke-gravity,sedov-gravity}. Despite this, the net circulation along the ring remains zero, with the flow on one side compensating the flow on the other. However, when the flow in one part of the ring is suppressed by the presence of a barrier, the steady state  which is created as a result exhibits a non-zero net circulation. To create a tunable barrier, we employ a second non-resonant laser to locally blueshift the polariton potential. This all-optical modulation is proportional to the intensity of the second laser and results from the repulsion between polaritons and the exciton reservoir cloud.
The schematic of such tilted polariton ring with an imbalance of flows along the two arms is shown in Fig.~\ref{fig2}{\bf a}, along with an electric-circuit analog in Fig.~\ref{fig2}{\bf b}.
By varying the position of the optically-created barrier and changing its power, we span a large range of circulations from negative to positive values. Fig.~\ref{fig2}{\bf c}--{\bf d} show examples of experimental density and phase profiles created on the polariton ring when the barrier is placed on the left ({\bf c}) and on the right ({\bf d}) sides of the ring.

In a steady state, the order parameter of the condensate must have the form $\psi(\theta) = \sqrt{\rho(\theta)}e^{i\phi(\theta)}$, where $\rho$ is the superfluid density and $\phi$ is the phase. The average imposed circulation $\langle k\rangle$ can be calculated for each given realisation by decomposing the experimental profiles $\psi(\theta)$ into plane waves as:
\begin{equation}\label{<k>}
\langle k\rangle = \sum |W_k|^2k, \qquad W_k\sim\int\limits_0^{2\pi}\!\sqrt{\rho(\theta)}e^{i\phi(\theta)}e^{-ik\theta}d\theta.
\end{equation}
Since the variable $\theta$ represents azimuthal angle, $k$ are dimensionless and represent the angular momentum eigenstate numbers in such a decomposition. In  Fig.~\ref{fig2}{\bf e}--{\bf f} we show examples for specific realisations (those given in Fig.~\ref{fig2}{\bf c}--{\bf d}) and the corresponding average externally-imposed circulations $\langle k\rangle$.
They change from realisation to realisation.
\vor{More examples of experimental profiles for various barrier positions and powers, and the corresponding decompositions~(\ref{<k>}) are provided in Supplementary Fig.~3.}\\

\begin{figure}[b]
\centering
\includegraphics[width=\textwidth]{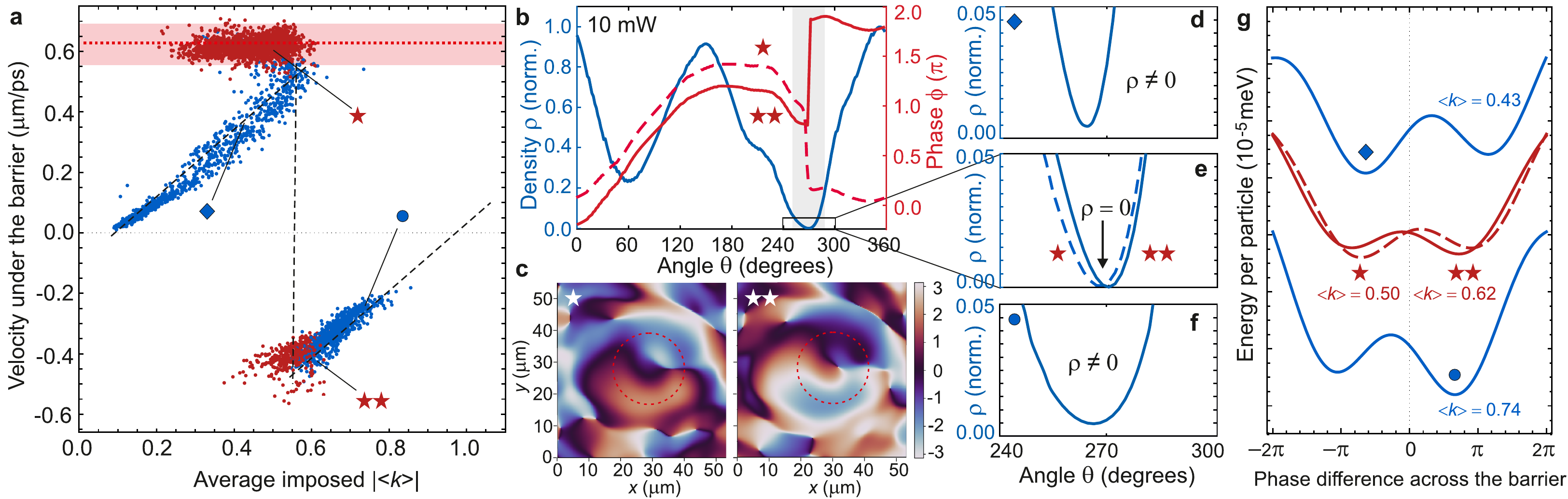}
\caption{\small\linespread{0.8} {\bf Two regimes of the weak link and the system's free energy landscapes.}
(a) The under-barrier velocity estimated from the phase jump across the barrier, $v\approx (\hbar/m)\Delta\phi/d$, versus the absolute value of imposed circulation along the ring. Here $m=0.5\times10^{-4}m_0$ is the effective polariton mass (in units of the free electron mass $m_0$). The data is united for all $\langle k\rangle$ obtained from both the left-placed and right-placed barriers; in order to have the velocities of the same sign, here for barriers on the left we defined $\Delta\phi=\phi(\theta_2)-\phi(\theta_1)$. All data points on this graph correspond to different experimental realisations, with colors of the points indicating the two essentially different regimes of the weak link operation: hydrodynamic (blue) and Josephson (red). The black dashed line provided by the theory is a guide to the eye for velocity behavior in a phase slip event. The red dotted line marks the \vor{average} sound velocity $c_{\rm s}$.
The red shaded range of velocities indicates the minimum and maximum values of $c_{\rm s}$, estimated from variations in condensate density at different barrier powers across the entire dataset (see SI).
(b) and (c) show two realisations in the Josephson regime, obtained for the right-placed barrier of power 10~mW, one of which shows zero winding (the dashed red line in {\bf b} and the left phase map in {\bf c}) and the other displaying winding 1 (the solid red line in {\bf b} and the right phase map in {\bf c}). The density distributions along the ring for those two realisations are almost coinciding. The gray-shaded area shows the barrier region. (d--f) Enlarged views of the density in the under-barrier region for realisations corresponding to different regimes. In particular, (e) is a zoom of (b), with the two lines corresponding to the realisations with zero (dashed) and non-zero (solid) windings. For both lines, the density touches zero under the barrier creating the condition for a phase jump. (d) and (f) show similar regions for the density profiles taken in the hydrodynamic regime at low and high $\langle k\rangle$ (as marked by the rhombus and circle symbols). In this regime, one sees that the density under the barrier does not go to zero and the wavefunction stays connected. (g) The system's free energy according to Eq.~(\ref{energy_theta}) with the experimental values of $\langle k\rangle$, $L$ and $E_{\rm kin}/E_{\rm tun}$ (see the SI) for selected realisations as marked, revealing the tilting and shifting washboard-shaped potentials. For small imposed circulations, the global minimum is that corresponding to no winding (top blue). For the barrier realisations resulting in $|\langle k\rangle|$ in the critical range (see the red points in {\bf a}), repetition of the same-setting experiment reveals that some realisations provide the minimum close to $\Delta\phi=-\pi$ (the dashed dark-red line, corresponding to $w=0$) and others at $\Delta\phi=\pi$ (the solid dark-red line, $w=1$), forming statistics among realisations. At the further increase of $\langle k\rangle$, the minimum at $\Delta\phi=\pi$ corresponding to $w=1$ is prevailing at all times (bottom blue).
}
\label{fig3}
\end{figure}

\noindent\textbf{Two regimes of the polariton weak link.} The external circulation along the free part of the ring creates a phase difference $\Delta\phi$ across the barrier region, resulting in a velocity of superflow under the barrier %that can be estimated as
$v\approx(\hbar/m)\Delta\phi/d$, where $m$ is the effective mass of the particles.
By increasing the barrier power, we increase the absolute value of $\langle k\rangle$ and---extracting both the phase difference and the barrier width from experimental profiles---we are able to track the under-barrier flow velocity. %which we plot in Fig.~\ref{fig3}{\bf a}.
For each experimental realisation, the phase difference and imposed current are evaluated individually, revealing the velocity-current relationship shown in Fig.~\ref{fig3}{\bf a} (it reports 4000 points corresponding to individual realisations, 500 per each barrier position and power). This comprehensive data set provides robust statistics on the current-phase relationship across the entire range of experimental repetitions, where we are able to span a range from very small to large values of circulation in both negative and positive domain.
Fig.~\ref{fig3}{\bf a} unites the data taken for different barrier positions and powers; in order to analyse the data in combination, for the right-placed barriers we plotted $v$ taking the phase difference in the opposite direction ($-\Delta\phi$).

For small powers, as long as the order parameter is not fully suppressed in the barrier region, the phase does not have room for ambiguity due to single-valuedness of $\psi$. This is the hydrodynamic regime, in which the phase difference across the barrier is bound to compensate for the flow along the opposite side of the ring. An example of the phase and density profiles in this regime is given in Fig.~\ref{fig2}{\bf c}.
However, when $|\langle k\rangle|$ is increased, we clearly see in Fig.~\ref{fig3}{\bf a} that the data points start deviating from the linear hydrodynamic dependence.
When the phase difference imposed on the barrier reaches $\pi$, within the repeated realisations of the same experiment we start to register stochastic sharp phase jumps in the weak-link region that reverse the direction of the under-barrier flow (see the red data points in Fig.~\ref{fig3}{\bf a}, which upon reaching a critical velocity experience a drop to negative values, creating two plateaus in an extended range of $\langle k\rangle$). Example of the density and phase profiles corresponding to this regime are shown in Fig.~\ref{fig3}{\bf b},{\bf c}: whereas the density distribution along the ring stays practically unchanged, the phase profiles can randomly show the absence or presence of a non-zero winding, for the same position and same power of the barrier.
We find that the critical velocity of the under-barrier flow that is reached when the behaviour of the weak link is changing (see the red dotted line in Fig.~\ref{fig3}{\bf a}) is approximately equal to the sound velocity $c_{\rm s}=\sqrt{gn/m}\approx 0.63~\mu$m/ps,
with $n$ denoting the average condensate density, \vor{$m$ the effective mass} and $g$ the polariton interaction constant.
To account for fluctuations in condensate densities across different realizations and barrier powers (see SI), the sound velocity  in Fig.~\ref{fig3}{\bf a} is shown with a corresponding variability range (the red-shaded range of velocities). At further increase of $|\langle k\rangle|$ the system exhibits a phase slip: the phase difference created by the external circulation is decreased by $2\pi$. If we keep increasing the power of the barrier, the hydrodynamic regime is recovered (blue data points in the region of large $|\langle k\rangle|$ in Fig.~\ref{fig3}{\bf a}). At this point, the under-barrier superflow velocity starts to grow again linearly with $\Delta\phi$, but with a non-zero winding of phase around the ring ($w=\pm1$ for barriers on the right and left, respectively).

Since it is apparent from Fig.~\ref{fig3}{\bf a} that in the critical range of imposed circulations the dependence $v(\langle k\rangle)$ is multi-valued, in order to accurately distinguish between the situations when the weak link is hydrodynamic and when it realises a Josephson junction, we take a look at the behavior of the order parameter in the weak-link region. Fig.~\ref{fig3}{\bf d}--{\bf f} show such regions enlarged, for four realisations in different regimes that correspond to chosen points on the velocity panel {\bf a}. As predicted, in the hydrodynamic regime the wavefunction is connected everywhere (Fig.~\ref{fig3}{\bf d}), supporting the superflow along the ring circumference. When the critical velocity is reached, however, we observe that the density under the barrier goes to exact zero in one point (Fig.~\ref{fig3}{\bf e}), breaking the continuity and creating the condition for stochastic phase jumps (see SI for more details). The superfluid motion in the weak link region in this regime is suppressed. Importantly, as demonstrated in Fig.~\ref{fig3}{\bf f}, when the hydrodynamic regime is restored for a higher barrier power, we see the reappearance of a fully connected phase and non-zero order parameter.

An important feature of our experiments is that both the barrier and the ring condensate are all-optically generated by external lasers. Although fluctuations in the laser intensities lead to variations in the barrier width and polariton densities (see Supplementary Fig.~2), there is no associated error in the evaluation of individual points in Fig.~\ref{fig3}{\bf a}, as both the average $\langle k \rangle$ and the phase difference are extracted independently for each experimental point. The effect of these fluctuations is therefore visible in the distribution of the whole dataset shown in Fig.~\ref{fig3}{\bf a}: despite the spread of points, the current-phase relationship remains clearly visible. \\

\noindent\textbf{System's free energy landscapes.}
As Josephson pointed out for superconductors~\cite{josephson2}, for a system which is interrupted by a barrier, the free energy of the system
\begin{equation}\label{GPE_func}
\mathcal{E}[\psi] = \oint Rd\theta\left[-\frac{\hbar^2}{2mR^2}\psi^*\frac{d^2}{d\theta^2}\psi + \frac{g}{2}|\psi|^4 + (U(\theta) - \mu)|\psi|^2\right]
\end{equation}
contains a contribution from the barrier region, which depends on the phases of the order parameter $\psi$ by the two sides of the weak link and whose magnitude becomes greater as the barrier is made thinner. In Eq.~(\ref{GPE_func}), $\mu$ is the condensate chemical potential and the optically-created potential $U(\theta)$ is nonzero only in the region of the weak link.
\vor{We use an equilibrium, zero-temperature model, similar to that applied in superconductors and atomic gases. While the system is driven-dissipative, the long polariton lifetime (100 ps) allows the effect of dissipation to be neglected in our steady-state observations. As we show below, this model provides a good correspondence with the experiment.}
The line integral in Eq.~(\ref{GPE_func}) is to be taken around the red dashed line in e.g. Fig.~\ref{fig3}{\bf c}, which can be split without the loss of generality into the barrier region of the width $d$ and the remainder of the ring of the length $L=2\pi R-d$, such that $\oint = \int_L+\int_d$. The last term in this sum represents the part of the free energy from the barrier $E_{\rm tun}$ similar to the one described by Josephson.
Then, to gain insight into the role of external circulation $\langle k\rangle$ imposed on the polariton ring, we consider an approximation where along the free part of the ring the coherent condensate wave function has a uniform density and a linearly changing phase (including the own phase and the one from the imposed motion), whereas at the barrier the wave function experiences a phase jump $\Delta\phi$. % between the right and left sides.
These approximations allow us to built an intuitive, minimal model that---despite its simplicity---captures the essential fundamental features of Josephson physics. Deriving the associated energy landscape in this fashion, we get the free energy in terms of the phase difference $\Delta\phi$:
\begin{equation}\label{energy_theta}
\mathcal{E}(\Delta\phi) = E_{\rm kin}(\Delta\phi)^2  - E_{\rm tun}\cos(\Delta\phi + \langle k\rangle L/R) + \text{const},
\end{equation}
where $E_{\rm kin} = \hbar^2\rho_0/(2mL) = \hbar^2N/(2mL^2)$ denotes the amplitude of the parabolic term, with $\rho_0=N/L$ and $N$ the number of particles in the polariton BEC. The detailed derivation is provided in Methods.

In Eq.~(\ref{energy_theta}), $\Delta\phi$ can be treated as a coordinate and $\mathcal{E}(\Delta\phi)$ as the potential energy for the system's dynamics, so that in equilibrium it is forced to stabilize in a global minimum of $\mathcal{E}(\Delta\phi)$.
By changing the width and height of the barrier in $U(\theta)$ one can manipulate both the circulation $\langle k\rangle$ and the ratio of the amplitudes $E_{\rm kin}/E_{\rm tun}$.
This effectively results in the reshaping (tilting and shifting) of the free energy landscape as a function of the phase difference $\Delta\phi$ across the barrier. Note that the ratio $E_{\rm kin}/E_{\rm tun}$ extracted from the experiments falls within the correct parameter range to enable the interplay between the two terms in (\ref{energy_theta}) (see Supplementary Fig.~4{\bf a}).
In our experimental data, any specific realisation allows to assess $E_{\rm kin}/E_{\rm tun}$ by analytical fitting of the density tails in the barrier region, as well as to define the barrier width $d$ and average $\langle k\rangle$ according to Eq.~(\ref{<k>}).
The corresponding energy landscapes $\mathcal{E}(\Delta\phi)$ calculated per one particle (divided by $N$) for four chosen realisations belonging to different regimes are plotted in Fig.~\ref{fig3}{\bf g}. The calculation details are provided in the SI, \vor{together with more examples for both left and right barriers of different powers in Supplementary Fig.~4}. For small imposed circulations, the free-energy landscape (top blue line in Fig.~\ref{fig3}{\bf g}) shows that the system in the steady state has to stay in the global minimum of $\mathcal{E}$ at $\Delta\phi<0$, which is shifting with the growth of $\langle k\rangle$. When $\langle k\rangle L/R$ approaches $-\pi$ for the left-placed barriers (or $+\pi$ for the barriers on the right) and $E_{\rm kin}/E_{\rm tun}$ is assuming its maximal values, the two minima of $\mathcal{E}(\Delta\phi)$ become shallow and close in energy (up to being degenerate). Repeated realisations of the experiment in this regime result in the statistics revealing one minimum lower than the other in some of realisations, and vice versa in the others (see the dashed and solid red lines in Fig.~\ref{fig3}{\bf g}). The dependence of $\Delta\phi$ on $\langle k\rangle$ in this regime is flat, with the phase difference across the barrier being constant (with a small variance) in the vicinity of $\pm\pi$. Provided by these energy considerations, the possibility to realise two distinct values of the phase difference (equilibrate in one of the two degenerate minima of $\mathcal{E}$) stochastically requires the existence of at least one point under the barrier where $\rho=0$ so that the phase is undefined. This is confirmed by looking at the density behavior under the barrier, see Fig.~\ref{fig3}{\bf e}. When $\langle k\rangle$ is increased and the phase slip occurs, the minimum corresponding to a non-zero winding becomes deeper and prevails in all realisations (the bottom blue line in Fig.~\ref{fig3}{\bf g}).\\

\begin{figure}[t]
\centering
\includegraphics[width=\textwidth]{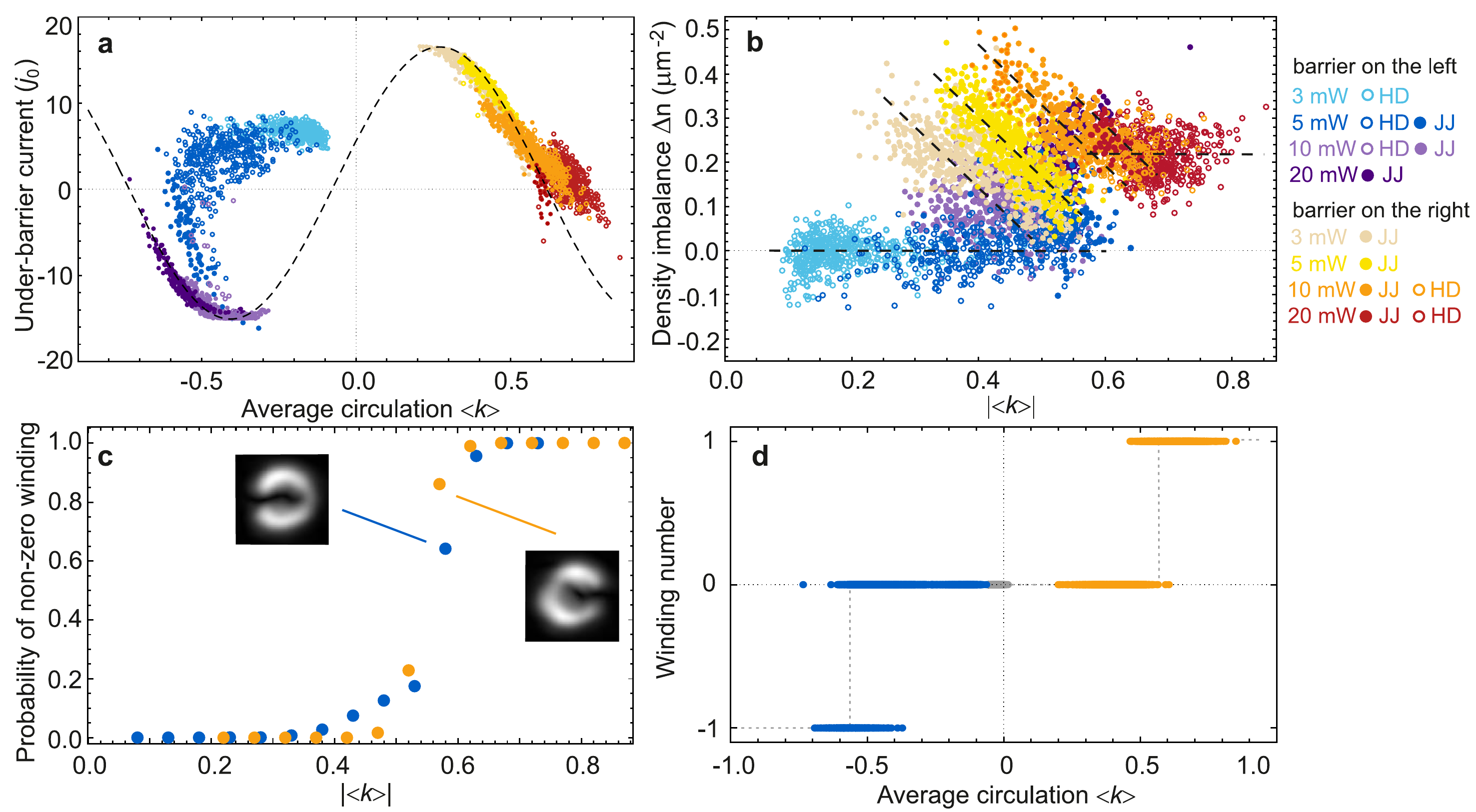}
\caption{\small\linespread{0.8} {\bf Josephson current and the density imbalance at the barrier.} (a) The under-barrier current $j_\theta = -j_c\sin{(\Delta\phi + \langle k\rangle L)}$ versus applied circulation $\langle k\rangle$, with $j_c=j_0\rho_{\rm 2D}/d$, $j_0 \lesssim 1$~ps$^{-1}$ given the polariton mass $m=0.5\times10^{-4}m_0$ (see Methods for calculation details), revealing the periodic change versus $\langle k\rangle$ for the Josephson regime. The points corresponding to the hydrodynamic regime (the open circles) do not follow the sinusoidal behavior (the black dashed line), being spread around zero.
(b) Density imbalance between the two sides of the barrier versus the average imposed circulation (absolute value). The offset due to the wedge for the non-interrupted ring is subtracted. The dashed black lines are guide to the eye indicating the independence of $\Delta n$ on $\langle k\rangle$ for the realisations in the hydrodynamic regime (open-circle points) and the linear change with $\langle k\rangle$ for the Josephson regime (full-circle points). The enlarged individual groups of points are provided in Suppl.~Fig.~7.
Colour code of the data points in panels {\bf a},~{\bf b} is chosen to distinguish different barrier positions and powers, as given in the legend.
(c) Probability of realisations with non-zero windings ($w=\pm1$) against the absolute value of average imposed circulation. (d) Winding numbers versus $\langle k\rangle$ for all realisations among all barriers. The color code in panels (b) and (d) is blue for the left- and yellow for the right-placed barriers. Gray points in (d) correspond to the case of an uninterrupted ring (without a barrier). The dashed line indicates the expected behaviour from our idealised theory for a fixed $L=7\pi R/4$.
}
\label{fig4}
\end{figure}

\noindent\textbf{Mass transport across the weak link and the tunneling current.} The Josephson regime is characterized by the undefined phase inside the weak link. The experimentally observed $\Delta\phi$ becomes independent of $-\langle k\rangle$ within a certain range, which results in the two plateaus of the red data points in Fig.~\ref{fig3}{\bf a}. One concludes that the observed phase difference, which stays fixed while the externally-imposed value $\langle k\rangle L/R$ is changing, contains a contribution from the inherent (tunneling) phase of the order parameter, $\Delta\tilde\phi$. The change of this tunneling phase by the two sides of the weak link is the origin of the Josephson (tunneling) current. Since there is no bulk superflow around the ring as soon as the density touches zero under the barrier, the only current through the weak link is the Josephson current %that is reversing its direction
$\propto\sin(\Delta\tilde\phi)$. Therefore, in this regime we expect the periodic change of the under-barrier current versus the imposed circulation $\langle k\rangle$, which is analogous to a supercurrent in a
SQUID becoming a periodic function of the magnetic flux piercing the ring~\cite{zimmerman}.
To retrieve the Josephson current, one needs to distinguish the tunneling contribution to the phase difference from the total measured $\Delta\phi$. To do so, we use the observed shapes of the wave function in the weak link region (their analytical fitting in the under-barrier region, see Supplementary Fig.~2 for details) and, assuming for the model case a linear growth of phase along the free part of the ring, make a rigid shift by the imposed phase difference $-\langle k\rangle L/R$ in the obtained functional dependence: $j_\theta = j_c\sin(\Delta\phi + \langle k\rangle L/R)$ which is provided in Methods. Plotting this way the under-barrier current against the imposed circulation $\langle k\rangle$ in Fig.~\ref{fig4}{\bf a}, we thereby confirm that the experimental points corresponding to the Josephson regime show periodic sinusoidal behavior, while the points corresponding to the hydrodynamic regime are scattered (due to the added contribution of bulk current).

Finally, we check the behavior of the density imbalance at the barrier versus the applied current in the two regimes. To do so, we extract from the density profiles in each experimental realisation the normalised imbalance $\Delta\rho=\rho(\theta_\vor{2})-\rho(\theta_\vor{1})$ for barriers on the right and the opposite for the left-placed barriers, and then multiply by the 2D density measured for each barrier power:
$\Delta n = \rho_{\rm 2D}\Delta\rho$. The density imbalance is plotted against the absolute value of the applied circulation in Fig.~\ref{fig4}{\bf b} for both positions of the barrier united. Individual groups of points for each power are provided separately in the SI. One can clearly see that in the realisations corresponding to the hydrodynamic regime there is no dependence of $\Delta n$ on $\langle k\rangle$. It suggests that in this regime the polaritons form a superfluid ring where the density imbalance is not dependent on the imposed flow with the mass transport sustained along the macroscopic flow path (ring circumference). When no winding of phase is present (light- and dark-blue open points), the average imbalance is zero, while for the hydrodynamic flow with $w=1$ (orange and red open points) the constant level of average $\Delta n$ is shifted, indicating that the superfluid density redistributes along the ring to sustain the winding. On the other hand, for the Josephson regime (solid points of all colors) there is a clear linear trend $\Delta n$ on $\langle k\rangle$  that indicates that as soon as the flow under the barrier is interrupted ($\rho=0$ in one point), there is no mass transport across the weak link and only tunneling is allowed.

To complete the analysis, we present the Josephson junction characteristics in terms of winding-current dependence. To do so, we unite the data of observed windings versus average imposed current $\langle k\rangle$ regardless of the barrier powers. Then we extract the statistical probability to realise a non-zero winding (in our case, $w=\pm1$) at each $|\langle k\rangle|$ within all 4000 frames (see the Supplementary Fig.~6 for a statistical histogram of the windings). Fig.~\ref{fig4}{\bf c} reveals that this probability experiences a sharp rise in the region of $|\langle k\rangle|\in(0.4,0.65)$, which corresponds to the imposed circulation values resulting in under-barrier velocities close to critical, in which the difference in phase of the weak link can equally be $\pm\pi$ (Josephson regime). In Fig.~\ref{fig4}{\bf d}, the phase winding around the ring is reported as a function of the average imposed current for each of the 4000 realisations, showing the jump between quantised levels corresponding to angular momentum $-1$, $0$ and $+1$.

In conclusion, we have demonstrated that the characteristic features of a weak link in a ring geometry can be measured in a photonic setting leveraging the nonlinear behavior of exciton-polaritons.
\vor{Even while the optical pumping introduces some level of experimental noise, the Josephson physics that we report here is extremely robust to such fluctuations. Notably, the current-density relation derived from the density and phase of the quantum fluid shows a quantitative correspondence with the minimalistic zero-temperature equilibrium model, formulated in terms of an energy functional and analogous to that of superconductors. This underlines that exciton-polariton platform, despite being inherently driven-dissipative, indeed belongs to the class of systems displaying the fundamental Josephson-junction behavior.}
While superconductor rings are crucial for high-precision measurements of magnetic fields (SQUIDs) or for generation of stable voltages (Josephson voltage standard), and superfluid helium or atomic BECs in a ring geometry have been proposed for the realisation of very sensitive gyroscopes~\cite{rotation},
the driven-dissipative nature of exciton-polaritons will allow for design of new schemes of operation based on fast nonlinear optics and coherent laser excitation.
Finally, despite this work having been %realised
performed at the temperature of 10K, suitable materials like transition-metal dichalcogenides~\cite{schneider,xiong} or perovskite crystals~\cite{su,bao} will make it possible to operate a polariton ring Josephson junction at room temperature.

\section*{METHODS}
\vspace{-12pt}
\noindent\textbf{Experimental methods.} The sample is a GaAs-based microcavity with 12 quantum wells embedded within two distributed Bragg reflector mirrors. The high-quality factor of the microcavity ($Q=10^5$) allows the attainment of extended polariton lifetimes (polariton linewidth $= 0.1$~meV) and it is designed to incorporate a significant wedge in the microcavity thickness of 1 $\mu$eV/$\mu$m. The Rabi coupling measures 15 meV, and the cavity-exciton detuning is $-2$~meV.

The formation of the ring polariton condensate occurs under non-resonant excitation using a continuous-wave laser tuned to 735 nm (polariton resonance at 774 nm), achieved by shaping the laser profile into a Mexican-hat intensity pattern (shown schematically in Fig.~\ref{fig1}{\bf a}) that serves both as the pumping beam and a confining potential.
The laser intensity is modulated using a phase-only spatial light modulator (SLM) positioned in the Fourier plane of the optical system. A conjugate gradient descent algorithm~\cite{conj_grad} combined with Gerchberg-Saxton iterations was used to optimize the formation of the Mexican hat in the image plane. A telescopic system is utilized to focus the beam on the sample surface and reduce the size of the Mexican hat by a factor of 50, resulting in a polariton ring condensate with the average radius $R=11.5$~microns (the inner and outer radii of the ring-shaped BEC are $R_1=8~\mu$m and $R_2=15~\mu$m, respectively), which is formed above the threshold density in the lowest confined mode of the potential.

The barrier used to split the annular polariton condensate consists of a second non-resonant continuous-wave laser beam tightly focused into a Gaussian spot with FWHM = 2 $\mu$m. The position of the second beam relative to the polariton ring condensate is adjusted using two external mirrors. The excitation power for the barrier laser is varied from 3~mW to 20~mW in each position.

The emission from the polariton ring condensate is collected in reflection configuration by the same imaging objective (NA = 0.5, focal length = 50 mm). A small portion of the condensate is selected in one arm of a Mach-Zehnder interferometer and used as the reference phase. The image of the ring condensate and the reference are combined at the exit of the Mach-Zehnder interferometer and imaged on the CCD detector to produce an interferogram. The phase and density information are then extracted from the interferogram using FFT as described in Ref.~\cite{dominici_ultrafast}.

The pump lasers are chopped at a frequency of 300 Hz with a duty cycle of 10\%, leaving an open time window of 333~$\mu$s. The detector is externally triggered to synchronise the acquisition with the opening of the chopper. Both the ring and barrier pump lasers are focused on the same point on the chopper plane to cut them simultaneously. For each pump power of the barrier and side of the ring (right or left), we take 500 realisations.\\

\noindent\textbf{Theoretical methods.} To describe our system we consider a one-dimensional (1D) geometry corresponding to a thin ring of an average radius $R=11.5~\mu$m, and assume all quantities dependent exclusively on the azimuthal angle $\theta$ (see Fig.~\ref{fig1}{\bf b}--{\bf d}). The validity of the 1D approximation as compared to the full 2D treatment is addressed in the SI.

To study the influence of external circulation $\langle k\rangle$ imposed on the polariton ring, in the order parameter $\psi$ in (\ref{GPE_func}) we explicitly separate the factor containing this imposed motion: $\psi \rightarrow \psi e^{i\langle k\rangle\theta} = \sqrt{\rho(\theta)}e^{i\tilde{\phi}(\theta)}e^{i\langle k\rangle\theta}$, with $\tilde{\phi}$ denoting the own phase of the condensate independent of the external circulation. The %toy
model consists of a simple consideration in which the coherent condensate wave function along most of the ring $\psi(\theta\in[\theta_{\rm R},\theta_{\rm L}])$  has a uniform density $\rho(\theta)=\rho_0$ and a linearly changing phase $\tilde{\phi}(\theta) + \langle k\rangle \theta = [\phi(\theta_\vor{1}) - \phi(\theta_\vor{2})]\theta R/L = -\Delta\phi \theta R/L$ (here $\theta_\vor{1,2}$ denote the positions of the \vor{two} sides of the barrier, so that $\theta_\vor{1}+d/R=\theta_\vor{2}$ and, given the periodicity, $\theta_\vor{2}+L/R=\theta_\vor{1}$). Having cut the barrier region out of the whole ring in this fashion, we can think of it as of a defect of the size $d$ on which the wave function $\psi_{\rm barrier}$ has a phase jump $\Delta\phi$ between the right and left sides. Then the barrier part of the energy functional (\ref{GPE_func}) can be integrated to produce a function $\propto -E_{\rm tun}\cos(\Delta\phi + \langle k\rangle L/R) + \text{const}$, with the tunneling energy $E_{\rm tun}$ defined by the specific shape of the density in the region. At the same time, substituting $\psi=\sqrt{\rho_0}\exp\{-i(\Delta\phi)\theta R/L\}$ to the ring part of Eq.~(\ref{GPE_func}), we get the free energy in terms of the phase difference $\Delta\phi$ given in Eq.~(\ref{energy_theta}) with
\begin{equation}\label{Etun}
E_{\rm tun} \approx 2N\,\frac{R}{L} \int\limits_{\substack{%
    \mspace{-9mu}\mathrm{barrier}\mspace{-9mu}\\
    \hidewidth\mathrm{region}\hidewidth
  }}\!\frac{[\mu-U(\theta)]d\theta}{1 + e^{d/\sigma} + e^{(\theta-\theta_\vor{1})R/\sigma} + e^{(\theta_\vor{2}-\theta)R/\sigma}},
\end{equation}
where $\sigma$ is the fitting parameter defined individually for each chosen experimental realisation. \vor{In Eq.~(\ref{Etun}), for the sake of analytical simplicity we assumed the slopes of the under-barrier wavefunction tails to be symmetric, which is not always the case. In the data analysis presented, those tails were fitted precisely with separate parameters $\sigma_1$ and $\sigma_2$.} The details of this calculation and additional considerations are provided in the Supplementary Information.\\

\noindent\textbf{Josephson current.} The analytical fitting of the experimental wavefunction profiles in the barrier region allows as well to calculate the expression for the under-barrier current, which in terms of the experimentally-measured phase difference reads $\boldsymbol{j} = (\hbar/m){\rm Im}\:\bigl[\psi^*\nabla\psi\bigr] = -\boldsymbol{j}_c\sin\Delta\tilde\phi = - \boldsymbol{j}_c\sin{(\Delta\phi + \langle k\rangle L/R)}$ with
\begin{equation}\label{j}
j_c = \frac{\hbar\rho_0}{m\sigma}\,\frac{2e^{d/\sigma}+e^{(\theta-\theta_\vor{1})R/\sigma}+e^{(\theta_\vor{2}-\theta)R/\sigma}}{(1 + e^{d/\sigma} + e^{(\theta-\theta_\vor{1})R/\sigma} + e^{(\theta_\vor{2}-\theta)R/\sigma})^2}\, \approx 0.1\,\frac{2\hbar\rho_0}{md} \approx 0.2\frac{\hbar}{m}\frac{\pi(R^2_2-R^2_1)}{L}\frac{\rho_{\rm 2D}}{d},
\end{equation}
where the number prefactor being an upper-bound estimate coming from the consideration that $\theta\in[\theta_\vor{1},\theta_\vor{2}]$, $\theta_\vor{2}-\theta_\vor{1}=d/R$. Additionally, to be able to plot the dependence of $j$ on $\langle k\rangle$ in Fig.~\ref{fig4}{\bf a} from the available experimental data, we roughly assumed $\sigma \sim d/2$ (more accurately, $\sigma$ depends not only on the width of $U(\theta)$, but also on its steepness and the polariton nonlinearity, i.e. it changes with the healing length).

\vspace{-10pt}
\section*{DATA AVAILABILITY}
\vspace{-12pt}
\noindent The data that support the findings of this study are available upon reasonable request. Correspondence and requests for materials should be addressed to D. Ballarini.

\vspace{-10pt}
\section*{ACKNOWLEDGEMENTS}
\vspace{-12pt}
 \noindent N.V. and A.G. who conducted the theoretical part of the work are funded by the Russian Science Foundation grant No. 24--22--00426 (\url{https://rscf.ru/en/project/24-22-00426/}). R.P., D.T., M.D.G., D.S. and D.B. acknowledge the financial support of the Joint Bilateral Agreement CNR--RFBR---Triennal Program 2021--2023, the MAECI project ‘Novel photonic platform for neuromorphic computing’, Joint Bilateral Project Italia--Polonia 2022--2023; European Union NextGenerationEU --- PNRR project, I-PHOQS Infrastructure (IR0000016, ID D2B8D520, CUP B53C22001750006); HORIZON--EIC--2023  PATHFINDER CHALLENGES --- European Innovation Council, Q-ONE (grant agreement no. 101115575); HORIZON--EIC--2023 PATHFINDER OPEN --- European Innovation Council, PolArt (grant agreement no. 101130304); PNRR MUR project: `National Quantum Science and Technology Institute' --- NQSTI (PE0000023).

\vspace{-10pt}
\section*{AUTHOR CONTRIBUTIONS}
\vspace{-12pt}
\noindent D.S. and D.B. coordinated the research project; R.P. and D.T. performed the measurements; N.V. and A.G. analysed the experimental data and developed the theory; growth was performed by K.B. and L.P.; N.V., D.S. and D.B. wrote the manuscript; all authors were involved in the discussion of results and the final manuscript editing.

\vspace{-10pt}
\section*{COMPETING INTERESTS}
\vspace{-12pt}
\noindent
The authors declare no competing interests.

%\newpage
%\section*{FIGURES}

\end{document}

\title{Exciton-polariton ring Josephson junction\\
Supplementary Information}

\author{Nina~Voronova}
%\email{neenoune@gmail.com}
\affiliation{National Research Nuclear University MEPhI (Moscow Engineering Physics Institute), Kashirskoe shosse 31, 115409 Moscow, Russia}
\affiliation{Russian Quantum Center, Skolkovo IC, Bolshoy boulevard 30 bld. 1, 121205 Moscow, Russia}

\author{Anna~Grudinina}
\affiliation{National Research Nuclear University MEPhI (Moscow Engineering Physics Institute), Kashirskoe shosse 31, 115409 Moscow, Russia}
\affiliation{Russian Quantum Center, Skolkovo IC, Bolshoy boulevard 30 bld. 1, 121205 Moscow, Russia}

\author{Riccardo Panico}
\email[Current affiliation: Institut f\"{u}r Angewandte Physik, Universit\"{a}t Bonn, Wegelerstra\ss e 8, 53115 Bonn, Germany]{}
\affiliation{CNR Nanotec, Institute of Nanotechnology, via Monteroni, 73100, Lecce, Italy}

\author{Dimitris~Trypogeorgos}
\affiliation{CNR Nanotec, Institute of Nanotechnology, via Monteroni, 73100, Lecce, Italy}

\author{Milena~De Giorgi}
\affiliation{CNR Nanotec, Institute of Nanotechnology, via Monteroni, 73100, Lecce, Italy}

\author{Kirk~Baldwin}
\affiliation{PRISM, Princeton Institute for the Science and Technology of Materials, Princeton University, Princeton, New Jersey 08540, USA}

\author{Loren~Pfeiffer}
\affiliation{PRISM, Princeton Institute for the Science and Technology of Materials, Princeton University, Princeton, New Jersey 08540, USA}

\author{Daniele~Sanvitto}
\email{daniele.sanvitto@nanotec.cnr.it}
\affiliation{CNR Nanotec, Institute of Nanotechnology, via Monteroni, 73100, Lecce, Italy}

\author{Dario~Ballarini}
%\email{dario.ballarini@nanotec.cnr.it}
\affiliation{CNR Nanotec, Institute of Nanotechnology, via Monteroni, 73100, Lecce, Italy}

\begin{abstract}
In this Supplementary Information file, we present the schematic of the experimental setup used to obtain the results presented in the main text, discuss the applicability of the 1D approximation based on the 2D experimental phase maps, and describe the fitting procedure for the experimental under-barrier density tails that allow to retrieve the tunneling energy and the amplitude of the Josephson current from experiment to be substituted into the model. We plot the distributions of angular momenta and the system's energy landscape for a set of specific realisations for the left- and right-placed barriers of different powers. Furthermore, we provide details of the calculation of the density imbalance dependence on the imposed flow and the enlarged plots of the dependencies presented in the main text. Extended (complementary) data on the condensate blueshift and the barrier widths versus the barrier power is also provided.
\end{abstract}

\maketitle

\begin{figure}[b]
    \centering
    \includegraphics[width=0.7\textwidth]{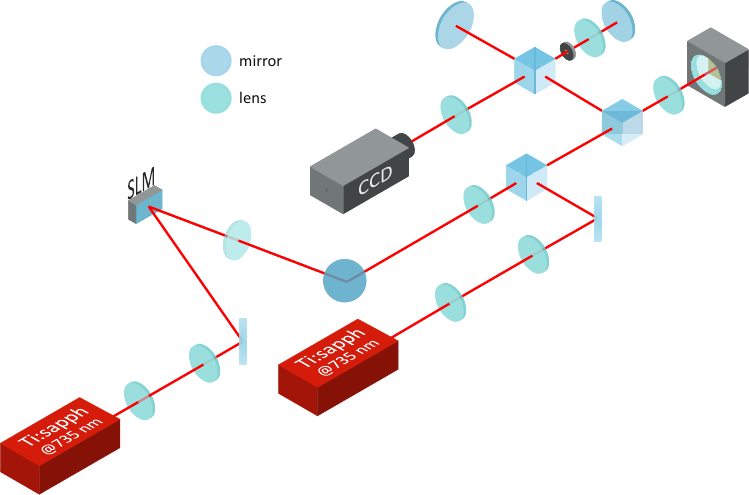}
    \caption{\small\linespread{0.8 }{\bf Experimental setup.} Two different laser sources are used to form the Mexican-hat potential and to form the potential barrier that splits the polariton ring-shaped condensate. Both lasers are tuned out of resonance with respect to the polariton energy. The laser forming the Mexican-hat potential serves as well to excite a large population of excitons that feed the polariton condensate. After collimation and expansion of the pumping laser, the beam is directed onto the SLM. Three lenses are used to direct the beam on the sample and form the image of the Mexican hat. The second laser beam is combined with the first laser beam through a beam splitter to form the potential barrier. The signal is collected in reflection configuration and another beam splitter is used to direct the signal into a Mach-Zehnder interferometer. In one arm of the interferometer, an iris is used to select a small portion of the condensate signal to be used as a reference phase. The interferogram produced by the image of the polariton ring condensate and the reference phase is recorded by a CCD.}
    \label{sfig1}
\end{figure}

\section{Microcavity wedge}

The primary effect of the microcavity wedge is to accelerate polaritons along the wedge direction, creating an effective gravity potential (see e.g.~\cite{snoke,nat578}). 
Since the photons inside a microcavity are confined by two mirrors in the $z$--direction, the corresponding component of the photon wavevector is quantized as $k_z = \pi n/L$, so that their energy offset and effective mass are defined by the cavity width $L$, as follows:
$$E_{\rm ph}(\boldsymbol{k}) = \hbar c k = \hbar c\sqrt{k_z^2 + k_\|^2} \approx \frac{\hbar c \pi n}{L} + \frac{\hbar^2k_\|^2}{2 m_{\rm ph}}, \qquad m_{\rm ph} = \frac{\hbar\pi n}{Lc},$$
where $c$ is the velocity of light in the medium, $k_\|$ is the in-plane wavevector, and $n$ is a natural number. 
When there is a slight change of the cavity width $L(x) = L_0 + \alpha x$, the energy offset (contributing to the detuning between photons and excitons) becomes dependent on $x$ as well,
$$\frac{\hbar c \pi n}{L(x)} \approx \text{const} - \frac{\hbar c \pi n}{L_0^2}\, \alpha x.$$
The second term thus becomes an analog of gravity forcing the motion of particles in the $x$--direction. Microcavity polaritons inherit this effect from photons. 
One can picture it as polaritons rolling off an inclined plane.

Our sample displays the wedge of $1~\mu$eV/$\mu$m, giving a variation of about 10~$\mu$eV from top to bottom of the ring, which induces motion along the ring as described in the main text (see Fig.~1{\bf c--d}). The phase profile shows a linear growth and decrease when going around the ring in the clockwise direction (starting from the top), which indicates constant velocity $\boldsymbol{v} = (\hbar/m)\nabla\phi$ along both arms of the ring from top to bottom.

\section{Validity of the 1D approximation}

The simplest way to reduce the two-dimensional (2D) description of particles moving on a ring to a one-dimensional (1D) problem consists of writing down the Laplacian in polar coordinates $r$ and $\theta$ and then, setting the radius to constant $r=R$, disregarding all terms containing the derivatives with respect to $r$. This approach works well for scalar particles (such as free particles on a ring or particles in the presence of a uniform field), but fails for systems with spin-orbit interaction (SOI) which is momentum-dependent and hence mixes the motion and pseudospin degrees of freedom (see e.g.~\cite{mukhejee}).
A more elaborate approach consists of directly considering the confining potential $U(r)$ in the Hamiltonian that confines the polaritons on the ring in the radial direction. Such $U(r)$ is small or zero in the vicinity of $R$ and large outside this region, so that this confining energy is much larger than the interaction energy, SOI energy (if applicable), and the kinetic energy in the azimuthal direction. This allows to separate variables $\Psi(r,\theta) = \chi(r)\psi(\theta)$ and assume that all particles are in the lowest radial state $\chi_1(r)$, while the potential $U(r)$ is steep enough to ensure that none of the perturbations are able to mix the states of the radial Schr\"{o}dinger equation~\cite{meijer,kozin}.

In our case the excitation conditions of the polariton annular BEC are such that the particles excited do not possess any specific polarization. In this case the TE-TM splitting producing the SOI terms in the underlying Hamiltonian can be safely neglected. Thus the problem is reduced to setting the average radius $R$ to be taken in all the formulae and then rewriting the kinetic energy in the shape $-\hbar^2/(2mR^2) \, \partial^2/\partial\theta^2$ given in Eq.~(2) of the main text. The geometrical parameters of the system set by the pump are described above: the inner radius of the ring $R_1 = 8~\mu$m and the outer radius $R_2=15~\mu$m. A rigorous calculation assumes defining the potential in the radial direction $U(r)$ which can be taken in the shape of an infinite square well of the width $R_2-R_1$~\cite{mukhejee} or, alternatively, as a harmonic confinement~\cite{kozin}. In both of these cases $\langle r\rangle$ calculated in the lowest radial state $\chi_1(r)$ normalised to unity (while the azimuthal function $\psi(\theta)$ is normalised to the number of particles $N$) with a great accuracy coincides with the arithmetic mean value $R=(R_2+R_1)/2 = 11.5~\mu$m.
To confirm the equivalence of taking any radius within the polariton ring-shaped BEC $R_1<R<R_2$ to describe the observed phenomena, we check the correspondence of the 2D experimental maps of the density (intensity) and phase of the condensate wavefunction collected in photoluminescence (PL) and shown in Fig.~1{\bf b},{\bf c} of the main text to the 1D cuts of the density and phase taken at a given radius within the ring (shown in panel {\bf d} of the same Figure). Looking at the phase profile one can see clearly that the choice of $R$ would not really alter the profiles. The phase profile reveals a uniform supercurrent in opposite directions along the two parts of the ring with respect to the microcavity wedge, with the zero net current in the case of no barrier.

%\begin{figure}[b]
%    \centering
%    \includegraphics[width=\textwidth]{SFig2.png}
%    \caption{{\bf 2D experimental maps and corresponding 1D profiles}. (a--b) Experimental maps of intensity (a) and phase (b) of the condensate emission for the case of the barrier on the right, with the barrier power 10~mW, for a selected realisation with zero winding of phase. (c) The density (blue) and phase (red) profiles along the central radius (red dashed lines in {\bf a},{\bf b}). (d--f) Same for a selected realisation with phase winding 1. The colorscales of phase in (b) and (e) are showing the numbers {\it modulo} $2\pi$. %In the unwrapped phase in (f), the $2\pi$--jump at the angle $\theta\approx100^\circ$ is eliminated.
%    }
%    \label{sfig2}
%\end{figure}

To confirm our assumption also for the case of the split ring, we make the same comparison for Fig.~3{\bf b},{\bf c} of the main text for the barrier of the power 10~mW placed on the right-hand side of the ring (with respect to the top of the wedge). The 2D maps of phase reveal that while the overall dependence of the phase on $\theta$ may slightly shift from one $r$ to another, the point where the phase experiences a $\pi$--jump (corresponding to the point where $\rho=0$, see the main text), both for realisations without and with the winding (Fig.~3{\bf c}), always stays at the same azimuthal position. The variation of the dependence $\phi(\theta)$ with $R$ along the free part of the ring is due to the fact that the density is not uniform any longer, while the current along the free part of the ring is fixed since it is dictated by the wedge. Thus we conclude that even though the considered ring is not thin ($R_2-R_1\sim R$), the 1D approximation works very well. At the same time, in the theory presented in the main text all calculations are performed in angular variables, and the specific choice of $R$ is only responsible for the scaling of the lengths.

\section{Fitting of the tails and calculation of $E_{\rm tun}$. Extended data for \\ circulation decompositions and tilted washboard energy profiles}

Using integration by parts and the periodic boundary conditions in the free energy of the system introduced in Eq.~(2) of the main text, we get
\begin{equation}\label{GPE_func}
\mathcal{E}[\psi] = \oint Rd\theta\left[\frac{\hbar^2}{2mR^2}\left|\frac{d\psi}{d\theta}\right|^2 + \frac{g}{2}|\psi|^4 + (U(\theta) - \mu)|\psi|^2\right],
\end{equation}
where $U(\theta)$ is the potential created by the barrier, $\mu$ is the condensate chemical potential, $m$ is the polariton effective mass, and $g$ is their interaction constant (recalculated for the 1D geometry). %\vor{We note that since our observations are made in a steady state of a driven-dissipative system, here we are using the equilibrium, zero-temperature model akin to that of superconductors or atomic gases. Another approximation made after} 
Next, we divide the integral in Eq.~(\ref{GPE_func}) into the barrier part and the ring part and assume that the condensate wave function along the free part of the ring has a uniform density and a linearly changing phase $\psi(\theta) = \sqrt{\rho(\theta)}\exp\{i(\tilde\phi(\theta) + \langle k\rangle\theta)\} = \sqrt{\rho_0}\exp\{i(-\Delta\phi)\theta R/L\}$, with $\Delta\phi$ the experimentally-observed phase difference across the barrier located between the angular coordinates $\theta_1$ and $\theta_2$ (we keep in mind that $\theta+2\pi$ coincides with $\theta$). %\vor{These approximations allow us to built an intuitive, minimal model that---despite its simplicity---captures the essential fundamental features of Josephson physics.} 
The phase difference $\Delta\tilde\phi$ is a degree of freedom in our consideration, while the imposed motion via the average circulation $\langle k\rangle$ creates ``phase pressure'' at the two sides of the barrier. 

In the idealistic-model case such a definition would require
$\psi_{\rm barrier} = \sqrt{\rho_0}\,\Theta(\theta_1-\theta) + \sqrt{\rho_0} e^{i\Delta\tilde\phi}\,\Theta(\theta-\theta_2),$
where $\Theta(\theta)$ denotes the Heaviside step-function, $\Delta\tilde\phi = \Delta\phi+\langle k\rangle L/R$. To avoid working with discontinuous functions, however, and to achieve better correspondence with the experiment, we fit instead the experimental under-barrier tails using the sigmoid functions:
\begin{equation}\label{sigmoid}
\psi_{\rm barrier}(\theta) \approx \frac{\sqrt{\rho_0 }}{1 + \exp\{(\theta-\theta_1)R/\sigma_1\}} + \frac{\sqrt{\rho_0} \,e^{i\Delta\tilde\phi}}{1 + \exp\{(\theta_2-\theta)R/\sigma_2\}},
\end{equation}
where the coordinates $\theta_{1,2}$ and $\sigma_{1,2}$ are fitting parameters that allow to extract $d=(\theta_{\rm 2}-\theta_{\rm 1})R$ from the wavefunction profiles in each specific experimental realisation. The constant-density level $\rho_0$ in Eq.~(\ref{sigmoid}) is taken in our fitting in such a way that the resulting density profile along the ring is normalized to the same number of particles as the experimental profile $\rho(\theta)$. Examples of such fitting are presented in Supplementary Fig.~\ref{sfig3}{\bf a} and {\bf b} for the cases of weak and strong barriers, respectively. The large variation of $\rho(\theta)$ around the constant $\rho_0$ outside the barrier for high barrier powers (see e.g. Suppl.~Fig.~\ref{sfig3}{\bf b}) is discussed below. 
To the right-hand side of panels {\bf a},~{\bf b} in Suppl.~Fig.~\ref{sfig3}, we provide exemplary fitting parameters using the shape~(\ref{sigmoid}) for the under-barrier wavefunction in 10 randomly chosen realisations with different barriers (the profiles for these realisations are presented in Supplementary Figure~\ref{sfig3new}). 
We note that since the optically-created barrier does not have a well-defined boundary, %and, at the same time, the length scale of the sigmoid drop $\sigma$ depends also on the barrier height and the polariton nonlinearity, only a rough
we use our estimate for $d=(\theta_{\rm 2}-\theta_{\rm 1})R$ obtained from the fitting only as a width of the Gaussian barrier shape $U(\theta)$ in further calculations that concern the selected experimental frame. For plotting Fig.~3{\bf a} and Fig.~4 of the main text, where the data on $d$ and $L$ for all 4000 realisations is required, we use the barrier widths extracted experimentally with a lesser accuracy from the cross-cut of PL emission at the condensate blueshift level (see Suppl.~Fig.~\ref{sfig3}{\bf c}, where the errors are given by the fluctuations of the width within the 500 realisations of the same barrier power). 

\begin{figure}[t]
    \centering
    \includegraphics[width=\textwidth]{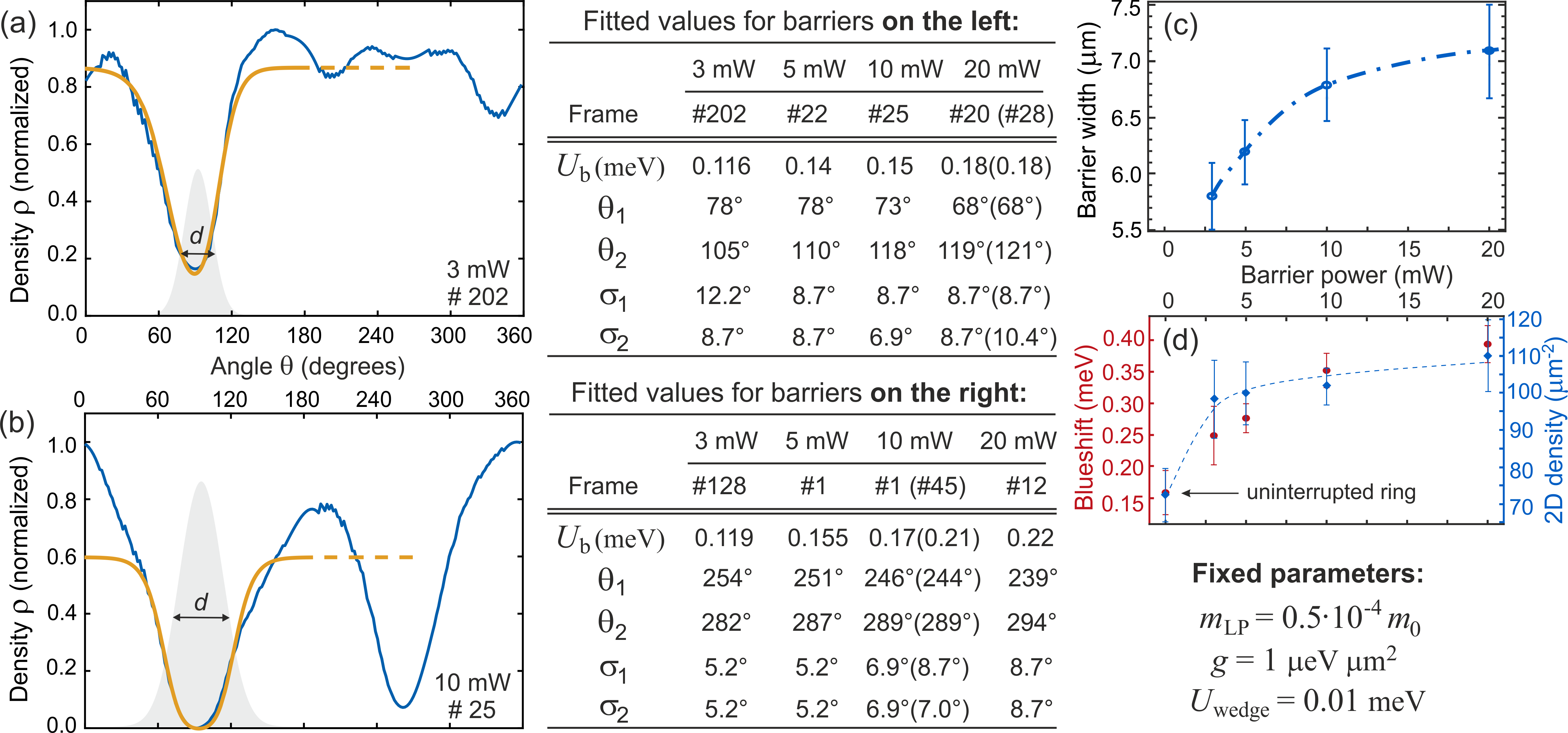}
    \caption{{\bf Details of the fitting, experimental blueshift, and barrier widths}. (a,b) The blue lines show the density profiles for the two exemplary experimental realisations, for the barrier on the left with the powers 3~mW (a) and 10~mW. The yellow lines show the sigmoid fitting of the density tails in the barrier region according to Eq.~(\ref{sigmoid}). In particular, for (a), the yellow line is given by $\psi_{\rm barrier} = 0.87[1/(\exp\{(\theta - 78)/12.2\}+1)+1/(\exp\{(105-\theta)/8.7\}+1)]$.
    For (b), $\psi_{\rm barrier} = 0.6[1/(\exp\{(\theta - 73)/8.7\}+1)+e^{i\pi}/(\exp\{(118-\theta)/6.9\}+1)]$. The gray shading illustrates schematically the potential of the barrier $U(\theta)$ as a Gaussian of FWHM equal $d$. The tables in the middle of the figure supply fitting parameters for ten specific experimental frames plotted in Suppl.~Fig.~\ref{sfig3new}. 
    (c) Experimental dependence of the average barrier width on the barrier power, with the error coming from fluctuations of $d$ among realisations. (d) Average experimental blueshift (red circles) counted on top of $E_0=1.603$~eV and 2D density (blue squares) versus the barrier power, with errors assessed from fluctuations among realisations. In the data used in the main text to show the presence of the two regimes, the average polariton density increases by $\sim$ 10\% when changing from the smallest to the largest barrier power. This small change is primarily a consequence of the reduced area of the region outside the barrier, that increases the polariton density for larger barrier width.}
    \label{sfig3}
\end{figure}

\begin{figure}[t]
    \centering
    \includegraphics[width=\textwidth]{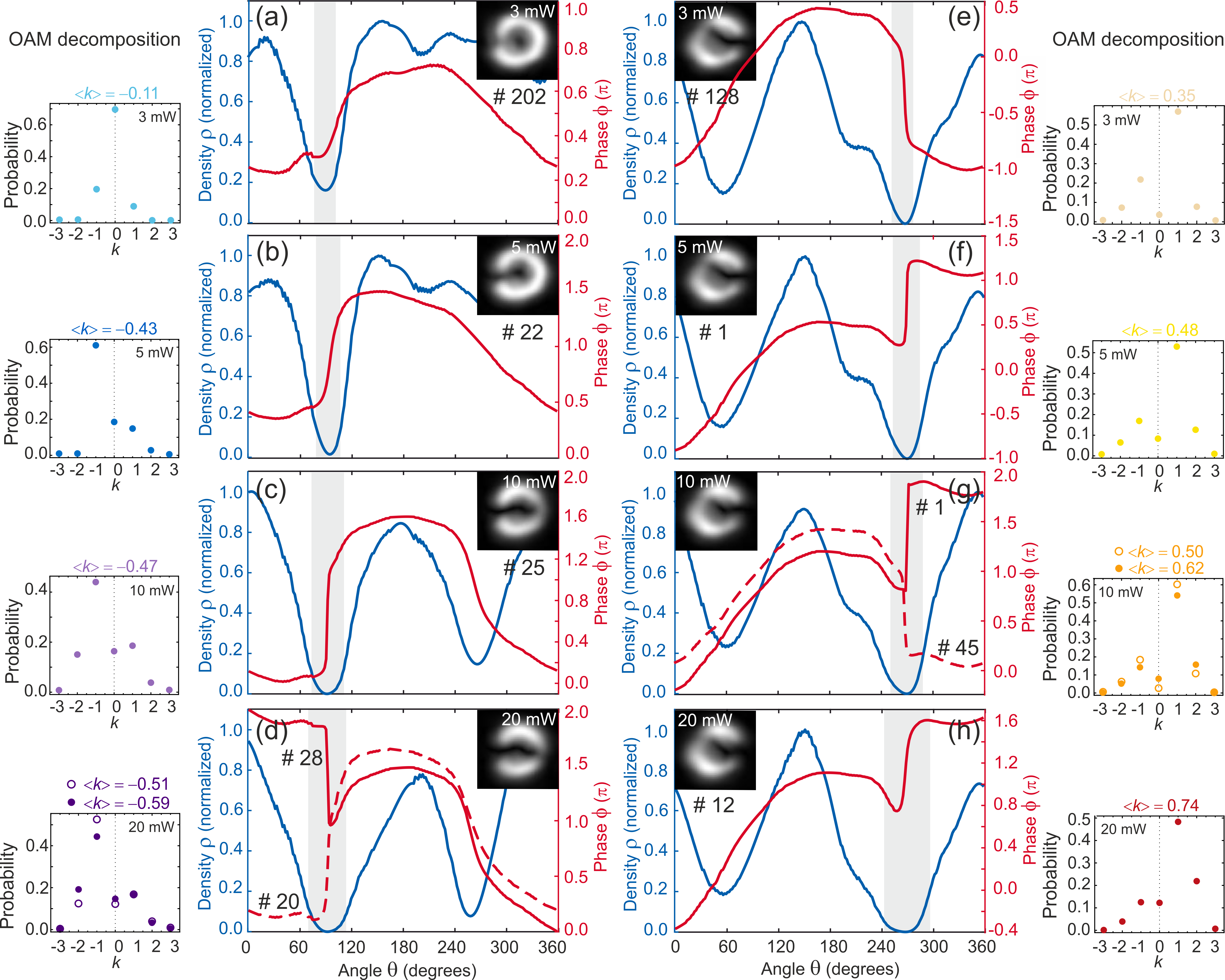}
    \caption{{\bf Additional exemplary experimental frames and corresponding $k$-distributions}  (a--d) for the barriers on the left and (e--h) on the right, powers as marked. Numbers of specific realisations are indicated on the panels. To the left and right of the two columns we present decompositions of corresponding realisations into plane waves [according to Eq.~(1) of the main text]. The color code corresponds to that used in Fig.~4 of the main text. In the panels (d) and (g), two realisations are plotted at the same time (with and without a winding). In corresponding momenta distributions, the open circles correspond to the realisations with $w=0$, while the full circles to $w=\pm1$ (the energy landscapes shown in Suppl.~Fig.~\ref{sfig2a}{\bf c},{\bf d} by the dashed and solid lines, respectively).}
    \label{sfig3new}
\end{figure}

The `barrier' part of the energy functional (\ref{GPE_func}) for can then be integrated with the substitution of Eq.~(\ref{sigmoid}) to produce a function of the phase jump across the barrier $\mathcal{E}[\psi_{\rm barrier}] \propto -E_{\rm tun}\cos(\Delta\phi+\langle k\rangle L/R) + \text{const}$, with the tunneling energy defined as
\begin{multline}\label{Ej}
E_{\rm tun} = 2\rho_0 \!\!\int\limits_{\substack{%
    \mspace{-9mu}\mathrm{barrier}\mspace{-9mu}\\
    \hidewidth\mathrm{region}\hidewidth
  }}\!\frac{Rd\theta}{1 + e^{d/\sigma} + e^{(\theta-\theta_{\rm 1})R/\sigma} + e^{(\theta_{\rm 2}-\theta)R/\sigma}}\! \left[\frac{\hbar^2}{2m\sigma^2} \frac{e^{d/\sigma}}{1 + e^{d/\sigma} + e^{(\theta-\theta_{\rm 1})R/\sigma} + e^{(\theta_{\rm 2}-\theta)R/\sigma}} \right. \\
\left. + \mu - U(\theta) - g\rho_0\left(\frac{1}{(1+e^{(\theta - \theta_{\rm 1})R/\sigma})^2} + \frac{1}{(1+e^{(\theta_{\rm 2} -\theta)R/\sigma})^2}\!\right)\! \right]\!,
\end{multline}
where for the sake of simplicity we assumed $\sigma_{\rm 1}\approx\sigma_{\rm 2}\equiv\sigma$, $U(\theta)$ is the Gaussian of the full width at half maximum (FWHM) equal to $d$, and $\mu$ is estimated from the experimental density (it is plotted in Suppl.~Fig.~\ref{sfig3}{\bf d} dependent on the barrier power). In $\mathcal{E}[\psi_{\rm barrier}]$, we have dropped the term $~\sim\cos^2\!\Delta\phi$, as it is proportional to the square of the under-barrier tails overlap and hence is much smaller than the term $\sim\cos\Delta\phi$. Given the experimental densities (see Suppl.~Fig.~\ref{sfig3}{\bf d}), we estimate $\mu \sim g\rho_0 \sim g_{\rm 2D}\rho_{\rm 2D}\sim 0.1$~meV. Hence the main contribution to $E_{\rm tun}$ is provided by the terms containing the overlap in the lowest power, which with the substitution $\rho_0=N/L$ is then reduced to
\begin{equation}\label{Etun}
E_{\rm tun} \approx 2N\,\frac{R}{L} \int\limits_{\substack{%
    \mspace{-9mu}\mathrm{barrier}\mspace{-9mu}\\
    \hidewidth\mathrm{region}\hidewidth
  }}\!\frac{[\mu-U(\theta)]d\theta}{1 + e^{d/\sigma} + e^{(\theta-\theta_{\rm 1})R/\sigma} + e^{(\theta_{\rm 2}-\theta)R/\sigma}}.
\end{equation}
We note that Eqs.~(\ref{Ej}) and (\ref{Etun}) present analytic shapes obtained with the simplifications described above. When processing the experimental data, for calculation of $E_{\rm tun}$ we used the full shape with $\sigma_1\neq\sigma_2$ (see the parameters in the Tables of Suppl.~Fig.~\ref{sfig3}) and without neglecting any contributions.

In Supplementary Fig.~\ref{sfig3new}, we plot ten specific realisations of the experiment with different barrier positions and powers, for which we estimated $E_{\rm tun}$ using the fitting (\ref{sigmoid}) described above. All panels are accompanied by the corresponding $k$-decompositions according to Eq.~(1) of the main text and the values of average imposed circulation $\langle k\rangle$. Calculating as well the kinetic term amplitude $E_{\rm kin}=\hbar^2\rho_0/(2mL)=\hbar^2N/(2mL^2)$ for each  realisation, we are able to follow the ratio $E_{\rm kin}/E_{\rm tun}$ against the value of average imposed circulation, as shown in Supplementary Fig.~\ref{sfig2a}{\bf a}. We note that for the parameters that are realised in our experiments, the ratio $E_{\rm kin}/E_{\rm tun}$ oscillates in the range $(0.02,0.15)$ that precisely corresponds to the values allowing the competition between the parabolic and cosine terms in the system's free energy $\mathcal{E}[\Delta\phi] = E_{\rm kin}(\Delta\phi)^2 - E_{\rm tun}\cos(\Delta\phi + \langle k\rangle L/R) + \text{const}$.

\begin{figure}[t]
    \centering
    \includegraphics[width=\textwidth]{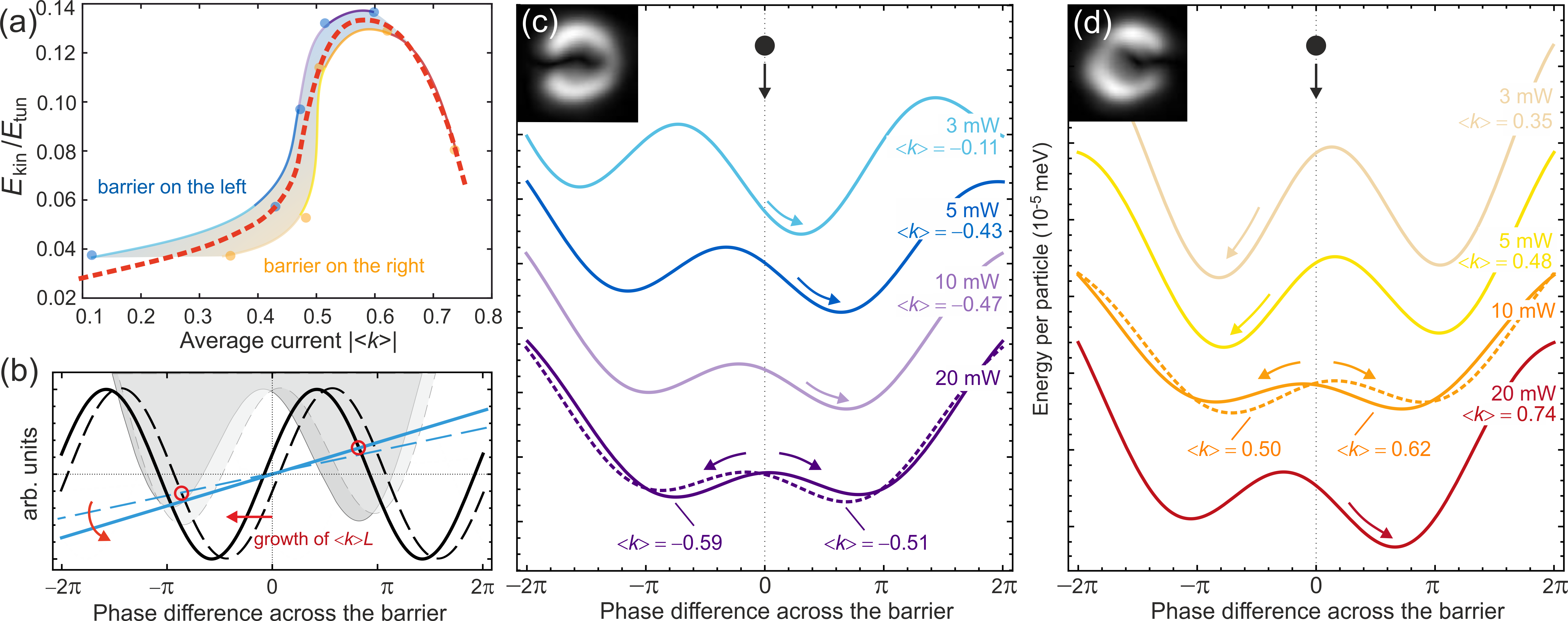}
    \caption{\small\linespread{0.8}
   {\bf The system's free energy landscape.} (a) The ratio $E_{\rm kin}/E_{\rm tun}$ extracted from selected experimental realisations for the barriers on the left (blue dots) and on the right (yellow dots) at different powers plotted versus imposed current $\langle k\rangle$ calculated for the same realisations. The red dashed line shows an average as a guide to the eye. (b) Graphical solution of the model Eq.~(4) of the main text, for the case of the barrier on the right, $E_{\rm kin}/E_{\rm tun} \approx 0.1$, dashed lines for $\langle k\rangle L/R=0.98\pi$, solid lines for $\langle k\rangle L/R=1.02\pi$, the red circles mark the values of $\Delta\phi$ corresponding to the global minimum of $\mathcal{E}(\Delta\phi)$ whose shape is shown by the gray-filled areas in the background. The red arrows indicate the direction of the shift of the lines with the growth of $\langle k\rangle$.  (c,d) The system's free energy (per particle) according to Eq.~(3) of the main text with the experimental values of $\langle k\rangle$, $L$ and $E_{\rm kin}/E_{\rm tun}$ for realisations from Suppl.~Fig.~\ref{sfig3new}, revealing the tilting and shifting washboard-shaped potentials. The corresponding imposed flows $\langle k\rangle$ due to wedge and the barrier powers are marked for all lines, while the $k$-distributions are plotted in Suppl.~Fig.~\ref{sfig3new}, with the same color code. The arrows indicate the position of the global minima. The insets show exemplary experimental PL profiles. In (c), for the left barrier power 20~mW, some realisations reveal the minimum close to $\Delta\phi=\pi$ (the dashed purple line, corresponding to $w=0$) and others at $\Delta\phi=-\pi$ (the solid purple line, $w=-1$), forming statistics among realisations. In (d), the same happens for the right barrier power 10~mW with the statistics between $\Delta\phi=-\pi$ (the dashed orange line, $w=0$) and $\Delta\phi=\pi$ (the solid orange line, $w=1)$, whereas at the further power increase to 20~mW, $\langle k\rangle L/R$ is larger than $\pi$ and the minimum at $\Delta\phi=\pi$ is prevailing among all realisations.}
    \label{sfig2a}
\end{figure}

Before analyzing the experimental data using our model, it is insightful to look for the position of the global minimum of $\mathcal{E}(\Delta\phi)$ by taking a derivative and tracking its zeros, which results in
\begin{equation}\label{derivative}
    \sin(\Delta\phi+\langle k\rangle L/R) = -2\frac{E_{\rm kin}}{E_{\rm tun}}\Delta\phi.
\end{equation}
Exemplary cases for a barrier on the right side of the ring, i.e. $\langle k\rangle>0$, are plotted in Suppl.~Fig.~\ref{sfig2a}{\bf b}. The black and blue lines represent the l.h.s. and r.h.s. of this equation, respectively, and the red mark encircles the intersection corresponding to the global minimum. 
The intersects the l.h.s. and r.h.s. of Eq.~(\ref{derivative}) furnish the discrete values of $\Delta\phi$ and of the corresponding circulating current, and they shift as $\langle k\rangle$ is increased (cf. the dashed and the solid lines in Suppl.~Fig.~\ref{sfig2a}{\bf b}). When $\langle k\rangle L/R$ overcomes $\pi$, a phase slip occurs, where $\Delta\phi$ corresponding to the global minimum of the free energy experiences a sudden jump from around $-\pi$ to a value in the vicinity of $+\pi$ (or the other way around for the case of left-placed barriers). 

When in the hydrodynamic regime at small $\langle k\rangle$, the flow along the free part of the ring is compensated by the flow under the barrier in the opposite direction (as in Fig.~2{\bf c} of the main text), resulting in zero winding of the phase $w=0$. In this case one can expand the l.h.s. of Eq.~(\ref{derivative}) around $\Delta\phi + \langle k\rangle L/R\approx0$ and obtain $\Delta\phi = -\langle k\rangle L/(1+2E_{\rm kin}/E_{\rm tun})$ which is a simple linear dependence. The corresponding linear part of the under-barrier velocity dependence on $\langle k\rangle$ is plotted in Fig.~3{\bf a} of the main text by the black dashed line. However as $\langle k\rangle$ is increased and $\langle k\rangle L/R$ overcomes $\pi$, it becomes energetically more favorable for the phase under the barrier to experience growth and produce the current under the barrier in the same direction as along the free part of the ring, resulting in $w=1$  (as in Fig.~2{\bf d} of the main text). This change describes a phase slip event, where $\Delta\phi$ corresponding to the global minimum of the free energy experiences a sudden jump by $2\pi$. 
This time $\Delta\phi+\langle k\rangle L/R\approx 2\pi$, so the Taylor expansion of the l.h.s. of Eq.~(\ref{derivative}) yields $\Delta\phi=(2\pi - \langle k\rangle L)/(1+2E_{\rm kin}/E_{\rm tun})$ (the black dashed line at $\langle k\rangle>0.6$ in Fig.~3{\bf a} of the main text). It is clear that the experimental blue points in the hydrodynamic regime in both cases follow the linear dependence predicted by this simple analysis.  
If the imposed circulation is further increased and $|\langle k\rangle L/R|$ goes over $2\pi$, $\Delta\phi$ and hence the velocity becomes positive again. The continuous tilting and shifting of the shape of $\mathcal{E}(\Delta\phi)$ would result in yet another phase slip of the global-minimum value of $\Delta\phi$ from around $-\pi$ to $+\pi$ at the moment when $w=2$ would be more favorable. In the Josephson regime which takes place before each phase slip when velocity reaches the critical value, the phase in the weak link region is no longer connected, hence the l.h.s. Eq.~(\ref{derivative}) cannot be expanded.

The circuit model depicted in Fig.~2{\bf b} of the main text presents a Josephson junction (JJ) biased by current $I_{\rm bias}$, which can be associated with a potential energy in the space of phase differences (at the two sides of the junction) $\Delta\phi$, known as the ``tilted washboard'' potential~\cite{qm_jj} in which the washboard piece $\sim\cos\Delta\phi$ is tilted by the linear term $\sim I_{\rm bias}\Delta\phi$. The motion of a point particle in this potential (in time) is a useful way to visualize the behavior of such a current-biased JJ, starting from a horizontal washboard ($I_{\rm bias} = 0$) and tilting it to one side as the current is applied. In our case, the energy landscape for the ``phase particle'' is given by the washboard combined with a parabolic kinetic term whose slopes and shift with respect to the cosine minima is controlled by the imposed circulation $\langle k\rangle$ along the free part of the ring. We do not look at the process in time; but since the system's excitation by default has no phase difference between any two points of the circumference, one may picture the process as dropping a ball at the coordinate $\Delta\phi=0$ into the potentials plotted in Suppl.~Fig.~\ref{sfig2a}{\bf c}--{\bf d} and looking at where it would end up, depending on the imposed flow applied to the system. These plots show $\mathcal{E}[\Delta\phi]$ per one particle (i.e. divided by $N$) for both left (c) and right (d) placed barriers for the realisations shown in Suppl.~Fig.~\ref{sfig3new} that span varying imposed circulations $\langle k\rangle$ both in negative and positive domain---supplementing those shown in Fig.~3{\bf g} of the main text. 

\begin{figure}[t]
    \centering
    \includegraphics[width=0.8\textwidth]{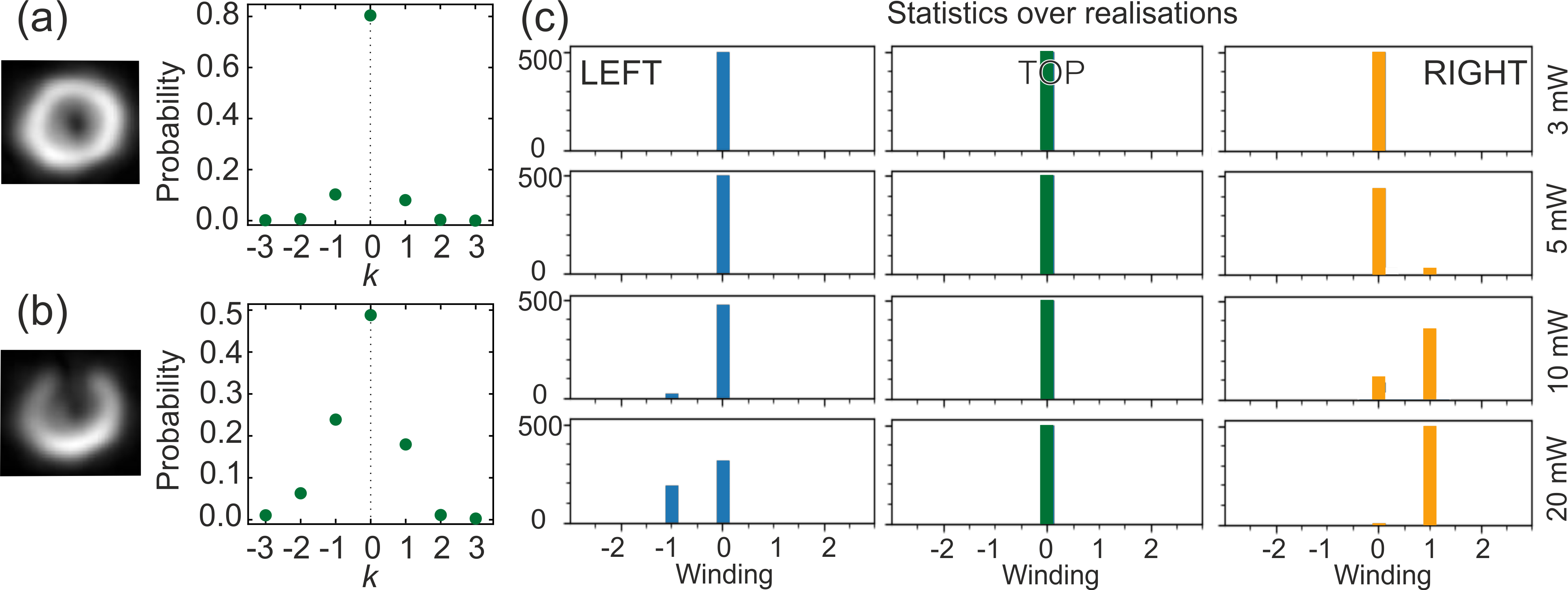}
    \caption{Examples of momenta distributions for the case of no barrier (a) and the barrier placed on the top of the ring with respect to the wedge (b). In both cases, the dependencies are symmetric around $k=0$. (c) Experimental statistics of phase windings around the ring collected for different barrier powers(as marked), for left-placed (the left column), top (in the middle), and right-placed (the right column) barriers. }
    \label{sfig4}
\end{figure}

The decompositions of the density and phase profiles $\sqrt{\rho(\theta)}e^{i\phi(\theta)}$ into plane waves [according to Eq.~(1) of the main text] are presented in Suppl.~Fig.~\ref{sfig3new}, to the left and right of the corresponding experimental frames, with the same color code. One can see that on top of the leading mode  $k=-1(+1)$ for all left (right) barrier realisations (except for the weak left-3mW barrier), there are always additional modes appearing on both sides of the main one, and their distribution around it is asymmetric. Generally the modes from $-2$ to $+2$ are present in all decompositions (except for the left-3mW case which is too weak to suppress the density enough), whose relative weights in the end define the average value $\langle k\rangle$. Superpositions of such counter-propagating waves on top of $k=\pm1$ result in the modulation of the density on top the uniform condensate moving around the ring, up to creating an additional minimum on the side opposite the barrier with the growth of its power (see Suppl.~Fig.~\ref{sfig3new}{\bf c--h}). 
Physically, the phase jump at the barrier locks the phases of the various components, allowing this density modulation to appear. It can be seen as the sum of two coherent counter-propagating currents or as the $p$--states of a confined wavefunction. The position of the minima in this case depends on the relative phase between the circulating currents, however in our case it becomes fixed because of the pinning of one of the two minima by externally-applied barrier.
This additional minimum in the density profile, however, would not appear in the case of zero net current, i.e. symmetric distribution of $k$ centered at zero (such situations are achieved for an uninterrupted ring or if the barrier is placed on top of the tilted ring, symmetrically with respect to the wedge, see Supplementary~Fig.~\ref{sfig4}{\bf a} and {\bf b}, respectively).

Suppl.~Fig.~\ref{sfig4}{\bf c} shows the collected statistics with respect to windings among all repetitions of the experiment for each barrier position and power, including the symmetric position of the barrier on top of the wedge (for comparison). Resulting statistics agrees with the theoretical description in terms of the energy landscape of the system shown in Suppl.~Fig.~\ref{sfig2a}{\bf c},{\bf d}: when one of the minima is global, one gets 100\% realisations with $w=0$ (for small powers) or $w=1$ (see the right-20~mW barrier); when the two minima of the energy become close in energy, the ``phase particle'' may end up in either of them, and both zero and nonzero windings become possible. In the case of the barrier on top, the average imposed circulation is zero, similarly to the case of no barrier at all (cf. panels {\bf a} and {\bf b} in Suppl.~Fig.~\ref{sfig4}), hence no phase slips and no windings appear for any barrier power.

\section{Zero of the under-barrier density and sorting of realisations into hydrodynamic and Josephson regimes}

As shown in the panels {\bf d--f} of Fig.~3 in the main text, we can distinguish between the hydrodynamic and Josephson regimes of the weak-link operation by looking at the behavior of the order parameter under the barrier. Mathematically, the condition is the disconnection of the wavefunction by the two sides of the barrier which is realised when the density goes to zero and the phase becomes undefined. Physically, the condition is the cease of the hydrodynamic current through the weak-link region and the stop of the mass transport. In our work, we verify both conditions.

To confirm that the density goes to zero, we check the behaviour the wavefunction (the raw unnormalized data). When the wavefunction passes through zero, i.e. has a node, it changes sign and hence the absolute value $|\psi|$ is broken (the derivative is discontinuous) and the tails by the two sides of the minimum are linear. When there is no node, $|\psi|$ around the minimum point is monotonous everywhere, with the derivative being continuous everywhere. 
In Supplementary Figure~\ref{sfig5new}, we present examples of such unnormalized wavefunction profiles. Panel (a) shows the comparison of realisations which belong to the hydrodynamic (blue) and Josephson (red) regimes, all realised within one set of experiments (for the left-placed 5 mW barrier). It is apparent that in the latter case the tails of $|\psi|$ linearize hence we can conclude, despite the discreteness of the experimental points, that indeed $|\psi|$ and therefore $\rho$ go through zero under the barrier. Panels (b) and (c) show one chosen example of such case for the wavefunction modulus and phase. Such analysis allows us to effectively distinguish between the two regimes for all the 4000 realisations for all barrier positions and powers, as is done in Fig.~3{\bf a} and Fig.~4{\bf a},~{\bf b} of the main text.

\begin{figure}[t]
    \centering
    \includegraphics[width=\textwidth]{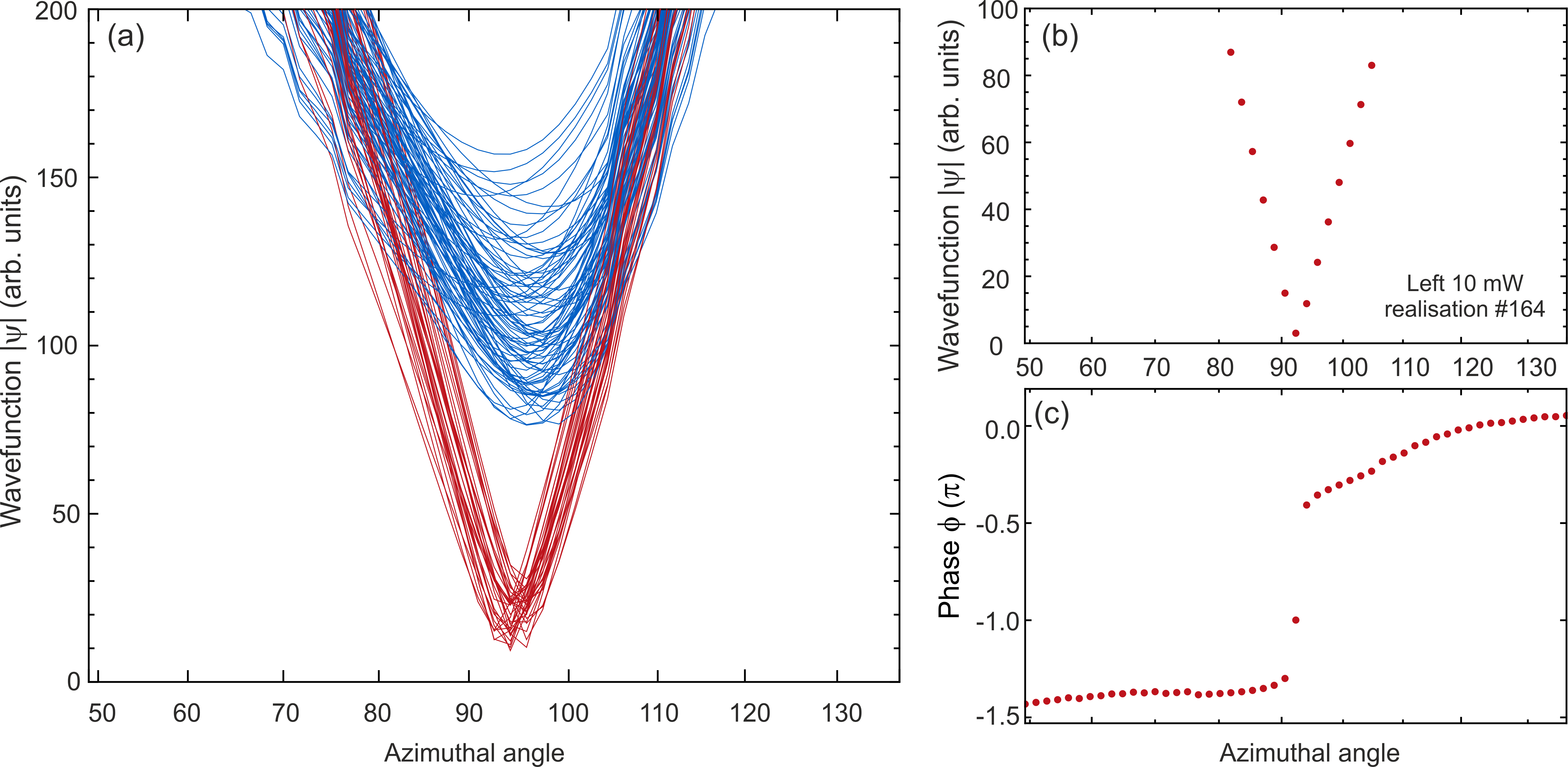}
    \caption{\small\linespread{0.8}
   {\bf Enlarged view of the unnormalized wave function under the barrier.} (a) Under-barrier wavefunction modulus for exemplary realisations revealing a node (linear shape of tails, red lines) in the Josephson regime in comparison with the parabolic shape (blue lines) in the hydrodynamic regime. (b--c) Enlarged view of one such realisation (left 10-mW barrier, frame No. 164) revealing a node of the wave function (b) and the corresponding phase jump by $\pi$ (c).}
    \label{sfig5new}
\end{figure}

%\section{Josephson current}

%The analytical fitting (\ref{sigmoid}) of the experimental wavefunction profiles in the barrier region allows to calculate the expression for the under-barrier current, which in terms of the experimentally-measured phase difference reads $\boldsymbol{j} = (\hbar/m){\rm Im}\:\bigl[\psi^*\nabla\psi\bigr] = -\boldsymbol{j}_c\sin\Delta\tilde\phi = - \boldsymbol{j}_c\sin{(\Delta\phi + \langle k\rangle L/R)}$ with
%\begin{equation}\label{j}
%j_c = \frac{\hbar\rho_0}{m\sigma}\,\frac{2e^{d/\sigma}+e^{(\theta-\theta_{\rm L})R/\sigma}+e^{(\theta_{\rm R}-\theta)R/\sigma}}{(1 + e^{d/\sigma} + e^{(\theta-\theta_{\rm L})R/\sigma} + e^{(\theta_{\rm R}-\theta)R/\sigma})^2}\, \approx 0.1\,\frac{2\hbar\rho_0}{md} \approx 0.2\frac{\hbar}{m}\bigl(R_2-R_1\bigr)\frac{\rho_{\rm 2D}}{d},
%\end{equation}
%where the number prefactor being an upper-bound estimate coming from the consideration that $\theta\in[\theta_{\rm L},\theta_{\rm R}]$, $\theta_{\rm R}-\theta_{\rm L}=d/R$ and the width of the ring $R_2-R_1$ appears from the ratio of the 1D and 2D densities. Additionally, to be able to plot the dependence of $j$ on $\langle k\rangle$ in Fig.~4{\bf b} of the main text from the available experimental data, we roughly assumed $\sigma \sim d/2$ (more accurately, $\sigma$ depends not only on the width of $U(\theta)$, but also on its steepness and the polariton nonlinearity, i.e. it changes with the healing length).

\begin{figure}[t]
    \centering
    \includegraphics[width=\textwidth]{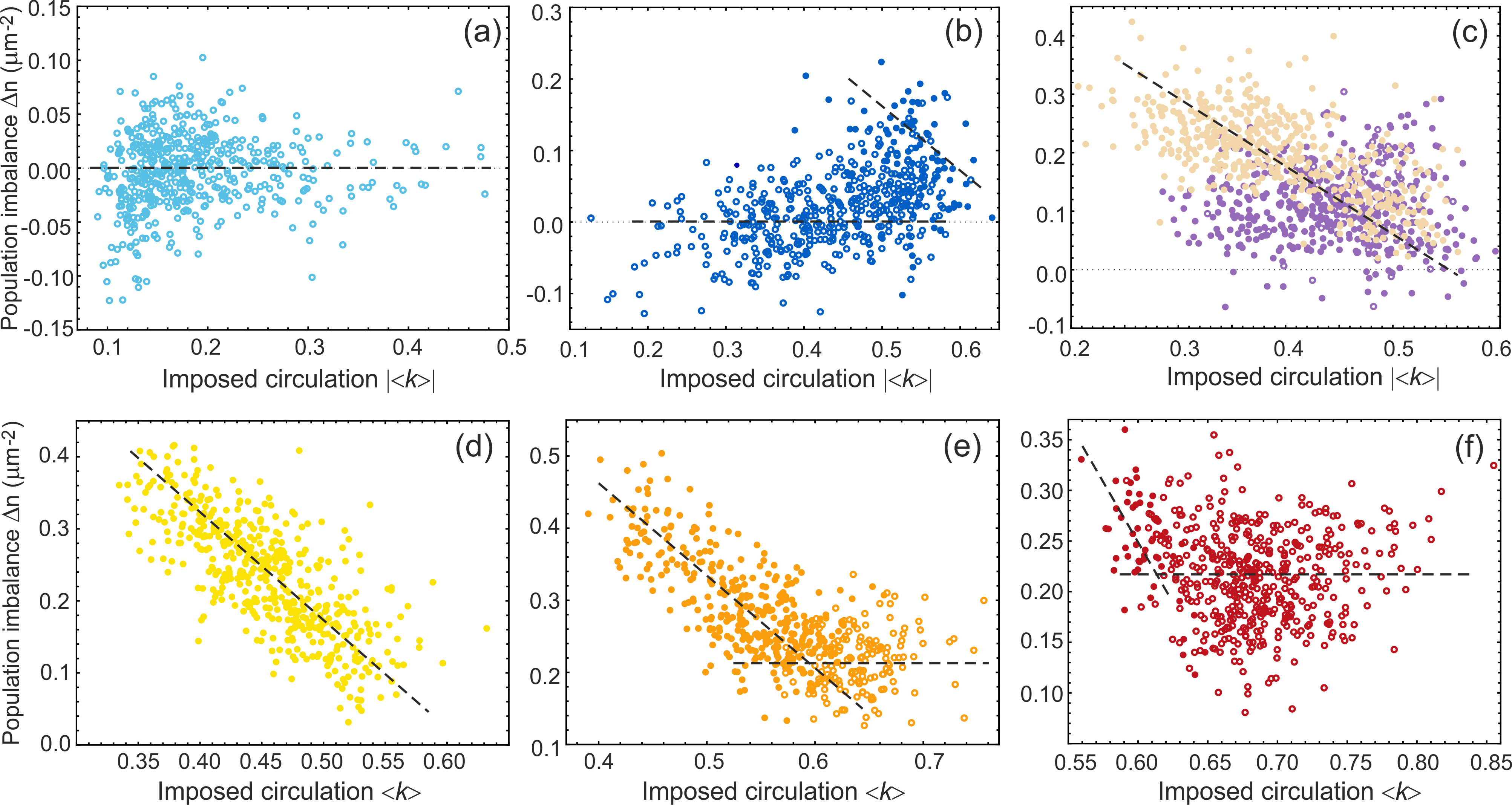}
    \caption{\small\linespread{0.8 }{\bf Density imbalance between the two sides of the barrier versus the applied flow.} Groups of points corresponding to different barrier positions and powers appearing in Fig.~4{\bf c} of the main text, plotted individually. (a) Left, 3~mW. (b) Left, 5~mW. (c) Left, 10~mW (lilac points) and right, 3~mW (beige points). (d) Right, 5~mW. (e) Right, 10~mW. (f) Right, 20~mW. The hydrodynamic regime with no dependence on $\langle k\rangle$ is apparent in {\bf a}, {\bf b}, {\bf e} and {\bf f} (open-circle points). The Josephson regime with the linear dependence on $\langle k\rangle$ appears partly in {\bf b}, in {\bf c}, {\bf d}, and partly in {\bf e}, {\bf f} (full points). The black dashed lines are guide to the eye showing the average constant level or the average slope of the linear dependence $\Delta n(\langle k\rangle)$.}
    \label{sfig5}
\end{figure}

%\section{Density imbalance}

Furthermore, we provide the enlarged plots of the density imbalance against the absolute value of the average imposed flow, presented in Fig.~4{\bf a} of the main text, and details of its retrieval from the measurements. This is another way to distinguish between the hydrodynamic and Josephson regimes, where in the former case there is no change in the imbalance when the imposed circulation $\langle k\rangle$ is growing, while in the latter case there is a clear linear dependence.

To calculate the imbalance for each experimental realisation we take the average barrier position by locating the minimum of density (it fluctuates around $\theta_{\rm barrier} = 90^\circ$ for the left barriers and around $\theta_{\rm barrier}=270^\circ$ for the right barriers) and track the difference between $\rho(\theta_{\rm 2})$ and $\rho(\theta_{\rm 1})$ where the positions $\theta_{\rm 2,1}=\theta_{\rm barrier}\pm d/2R$, with the average $d$ among the frames with the same power (see Suppl.~Fig.~\ref{sfig3}{\bf c}). Each density profile is normalised to unity before taking these values, and then multiplied by the 2D density corresponding to the given barrier power (see the densities in Suppl.~Fig.~\ref{sfig3}{\bf d}). %We note here that the growth of the average density that is observed with the increase of the barrier power is mainly due to the effective reduction of available space as the width of the barrier is growing. Our analysis revealed that there is no substantial feeding from the barrier itself. 
The resulting value $\Delta n = \rho_{\rm 2D}[\rho(\theta_{\rm 2})-\rho(\theta_{\rm 1})]$ is then plotted as a point on $\Delta n(\langle k\rangle)$ graph corresponding to the values of $\langle k\rangle$ calculated for the same experimental realisation. For the barriers placed on the left side of the ring (with respect to the wedge), $\langle k\rangle$ and $\Delta n$ calculated as described are negative, so we take absolute values to achieve the full picture presented in the main text in Fig.~4{\bf a}. Supplementary Figure~\ref{sfig5} shows the data points distribution for all experimental realisations for each barrier position and power. The large variance of the points around their average lines (shown by the dashed lines) originates from the fact that the barriers positions and widths in different experimental realisations slightly vary, while the calculation described above is done using a fixed barrier position $\theta_{\rm barrier}$ and width $d$. Nevertheless, the panels corresponding to different barrier powers clearly reveal the two trends characteristic for the connected-phase (hydrodynamic) and jumping-phase (Josephson) regimes of the weak link operation.

\quad\\